\documentclass[twocolumn]{aastex63}

\usepackage{amsmath}
\usepackage{amssymb}
\usepackage[inline]{enumitem}

\graphicspath{{./}{figures/}}

\accepted{to AJ February 29, 2020}

\shorttitle{WFC3/UVIS G280 grism data reduction and analysis}
\shortauthors{Wakeford et al.}

\begin{document}

\title{Into the UV: A precise transmission spectrum of HAT-P-41b using \\Hubble's WFC3/UVIS G280 grism}

\correspondingauthor{Hannah R. Wakeford}
\email{stellarplanet@gmail.com}

\author[0000-0003-4328-3867]{H.R. Wakeford}
\affil{School of Physics, University of Bristol, HH Wills Physics Laboratory, Tyndall Avenue, Bristol BS8 1TL, UK}
\affil{Space Telescope Science Institute, 3700 San Martin Drive, Baltimore, MD 21218, USA}

\author[0000-0001-6050-7645]{D.K. Sing}
\affil{Department of Earth \& Planetary Sciences, Johns Hopkins University, Baltimore, MD, USA}
\affil{Department of Physics \& Astronomy, Johns Hopkins University, Baltimore, MD, USA}

\author[0000-0002-7352-7941]{K.B. Stevenson}
\affil{JHU Applied Physics Laboratory, 11100 Johns Hopkins Rd, Laurel, MD 20723, USA}

\author[0000-0002-8507-1304]{N.K. Lewis}
\affil{Department of Astronomy and Carl Sagan Institute, Cornell University, 122 Sciences Drive, Ithaca, NY 14853, USA }

\author{N. Pirzkal}
\affil{Space Telescope Science Institute, 3700 San Martin Drive, Baltimore, MD 21218, USA}

\author[0000-0001-6352-9735]{T.J. Wilson}
\affil{University of Exeter, Physics Building, Stocker Road, Exeter, Devon, Ex4 4QL, UK}
\affil{Space Telescope Science Institute, 3700 San Martin Drive, Baltimore, MD 21218, USA}

\author[0000-0002-8515-7204]{J. Goyal}
\affil{Department of Astronomy and Carl Sagan Institute, Cornell University, 122 Sciences Drive, Ithaca, NY 14853, USA }

\author{T. Kataria}
\affil{NASA Jet Propulsion Laboratory, 4800 Oak Grove Drive, Pasadena, CA 91109, USA}

\author[0000-0001-5442-1300]{T. Mikal-Evans}
\affil{Kavli Institute for Astrophysics and Space Research, Massachusetts Institute of Technology, 77 Massachusetts Avenue, 37-241, Cambridge, MA 02139, USA}

\author{N. Nikolov}
\affil{Department of Physics \& Astronomy, Johns Hopkins University, Baltimore, MD, USA}

\author{J. Spake}
\affil{Division of Geological and Planetary Sciences, California Institute of Technology, Pasadena, CA 91125, USA}

\begin{abstract}
The ultraviolet-visible wavelength range holds critical spectral diagnostics for the chemistry and physics at work in planetary atmospheres. To date, exoplanet time-series atmospheric characterization studies have relied on several combinations of modes on Hubble's STIS/COS instruments to access this wavelength regime. Here for the first time, we apply the Hubble WFC3/UVIS G280 grism mode to obtain exoplanet spectroscopy from 200-800\,nm in a single observation. We test the G280 grism mode on the hot Jupiter HAT-P-41b over two consecutive transits to determine its viability for exoplanet atmospheric characterization. We obtain a broadband transit depth precision of 29--33\,ppm and a precision of on average 200\,ppm in 10\,nm spectroscopic bins. Spectral information from the G280 grism can be extracted from both the positive and negative first order spectra, resulting in a 60\% increase in the measurable flux. Additionally, the first HST orbit can be fully utilized in the time-series analysis. We present detailed extraction and reduction methods for use by future investigations with this mode, testing multiple techniques. We find the results fully consistent with STIS measurements of HAT-P-41b from 310--800\,nm, with the G280 results representing a more observationally efficient and precise spectrum. HAT-P-41b's transmission spectrum is best fit with a model with T$_{eq}$=2091\,K, high metallicity, and significant scattering and cloud opacity.  With these first of their kind observations, we demonstrate that WFC3/UVIS G280 is a powerful new tool to obtain UV-optical spectra of exoplanet atmospheres, adding to the UV legacy of Hubble and complementing future observations with the James Webb Space Telescope.
\end{abstract}


% ------------------------
% SECTION:
% INTRODUCTION
% ------------------------
\section{Introduction} \label{sec:intro}

The characterization of planetary atmospheres in the solar system and beyond has long leveraged the ultra-violet (UV) to near-infrared (NIR) spectroscopic capabilities of the Hubble Space Telescope (HST). Observations with HST have been critical in the exploration of the chemical composition, climate, and aerosol properties of exoplanet atmospheres \citep[see][and references therein]{kreidberg2018}. With the help of HST we now know that clouds and hazes are likely present in all types of exoplanetary atmospheres \citep[e.g.][]{marley2013,helling2019, Wakeford2019RNAAS}, but we currently lack information related to their abundances, physical properties and extent throughout the atmosphere. We also know that exoplanets exhibit extended upper atmospheres with evidence for atmospheric escape \citep[e.g.][]{ehrenreich2014,Bourrier2018,sing2019}, but struggle to connect physical processes in the lower and upper portions of exoplanet atmospheres. 

The UV through optical (200--800\,nm) spectra of planets hold rich information about the chemistry and physics at work across a broad range of atmospheric pressures. In the solar system, UV and near-UV spectroscopy has been critical in identifying and measuring the abundances of a variety of hydrocarbon and sulfur-bearing species, produced via photochemical mechanisms, as well as oxygen and ozone and more. For exoplanets, UV to near-UV spectroscopy has been especially useful for constraining aerosol properties and exploring atmospheric chemistry in hot ($>$1000\,K) atmospheres \citep[e.g.][]{sing2016_nature, Evans2018}. To date, only a handful of exoplanets have been probed in the critical 200--400\,nm wavelength range that crosses the optical to UV boundary.  Results from these studies have been mixed, limited by the wavelength coverage and sensitivity of the workhorse instrument for such studies, HST’s Space Telescope Imaging Spectrograph (STIS) G430L and E230M gratings. 

It is important to remember that none of HST's instruments or modes were specifically designed to support exoplanet observations. It has only been through the development of new observational strategies, such as spatial scanning \citep{mccullough2012, mccullough2014}, and data reduction techniques that the potential for HST to probe exoplanet atmospheres has been achieved. In general, slitless spectroscopic observing modes have been preferred for high-precision time-series observations of exoplanets that transit, pass in front of, their host stars because they typically offer more throughput and temporal stability. The slitless spectroscopy capabilities HST's Wide Field Camera 3 (WFC3) have been heavily used by the exoplanet community at infrared wavelengths (750--1800\,nm) with the G102 and G141 grisms. However, HST's WFC3 UV/Visible (UVIS) channel also offers slitless spectroscopy in the UV through visible (200--800\,nm) wavelength range that has yet to be leveraged for exoplanet observations. In fact, this mode has only been employed in a handful of scientific investigations, first used as part of HST WFC3 early release science programs in cycle 16 (2006), however, none of the G280 work was published from this study.

Here we detail for the first time the observations, spectral extraction, and analysis processes taken to apply Hubble's WFC3/UVIS G280 spectroscopic grism to transiting exoplanet observations. We first introduce the challenges in using the UVIS G280 grism in \S\ref{sec:UVIS}. In \S\ref{sec:obs} we detail the observations and spectral extraction procedures used. We then detail the broadband time-series analysis using two systematic reduction techniques in \S\ref{sec:analysis}. We use \textit{Spitzer} transit observations to refine system parameters and update the orbital ephemeris in \S\ref{sec:spitzer} and \ref{sec:o-c}. We outline the spectroscopic analysis in \S\ref{sec:spec_analysis} and discuss the results in \S\ref{sec:discussion} including searching for evidence of atmospheric escape and comparisons to STIS data. We then conclude with a summary of our results and the potential of WFC3/UVIS G280 for future exoplanet investigations. 

% ------------------------
% SECTION:
% INTRO TO UVIS
% ------------------------
\section{Introduction to the UVIS G280 grism}\label{sec:UVIS}
The WFC3 instrument on HST is fitted with two channels, UVIS and IR. Across these two channels are three slitless spectroscopic grisms: G280 in UVIS and G102 and G141 in the IR channel. The IR grisms have been extensively applied to exoplanet atmospheric studies with increasing success at the advent of spatial scanning \citep{mccullough2012}, where HST slews in the cross-dispersion direction to spread the target light over a column of pixels (e.g., \citealt{deming2013,kreidberg2014a,wakeford2013,wakeford2016,dewit2016}). However, the UVIS G280 grism has not had such usage despite large throughput in the near-UV (NUV) and wide coverage from 200--800\,nm. More commonly, studies that cover 300--900\,nm are conducted with multiple observations using HST's STIS G430L and G750L low resolution gratings from 300--550\,nm and 500--900\,nm respectively (e.g., \citealt{nikolov2014,sing2016_nature,Lothringer2018b}) despite their comparatively low throughput (Fig.\,\ref{fig:throughputs}). 

The UVIS grism, however, comes with several quirks that make it difficult to observe with and challenging to analyze. A number of these challenges will also affect observations with James Webb Space Telescope's (JWST) spectroscopic instrument modes. Therefore, WFC3/UVIS G280 is a current working example of the challenges that will be faced with JWST. Here we detail each of the challenges associated with WFC3's UVIS grism and also the advantages it has over other instrument modes in the NUV to optical wavelengths. 

% ------------------------
% Challenges
% ------------------------
\subsection{Challenges}
We detail some of the challenges encountered with this dataset and those expected in the general use of this instrument mode for exoplanet time-series characterisation. 

- \textbf{Curved spectral trace:} The trace for spectral order with the G280 grism is strongly curved at shorter wavelengths. The trace is best fit with a 6th order polynomial function detailed by \citet{Pirzkal2017_G280} and section \ref{sec:obs}. This curvature causes it to be offset in the cross-dispersion direction from the 0th order position, meaning subarray sizes need to be carefully chosen. Unlike the IR grisms, the spectra should not be spatially scanned as this would result in overlapping wavelength regions along the scan direction. 

The curved spectral trace also introduces a non-linear wavelength solution, meaning each pixel has a slightly different wavelength range than the surrounding pixels in that column. The wavelength solution is therefore extracted relative to the fitted trace position on the detector with a 6th order polynomial. 

- \textbf{Multiple overlapping orders:} Additional spectral orders, both positive and negative, overlap with the first order spectra at wavelengths greater than 550\,nm. In many cases these additional orders will be much dimmer than the first order spectrum and not impact the observations. However, for stars bright in the NUV such as O,B,A stars the additional spectral orders may impact the spectral extraction. 

In the presented case, the second order spectrum is $\approx$\,65$\times$ dimmer than the primary spectrum in both positive and negative orders. This would negligibly contribute to the measured transit depths causing the measured planetary radius ($R_{m}$) to be $\approx$\,99.24\% of the true planetary radius ($R$) following, 
\begin{equation}
\frac{R_{m}}{R} = \sqrt{\frac{1}{1+\frac{1}{65}}} = 0.9924.
\end{equation}

- \textbf{Geometric distortion:} Using the grism filters causes the spectra to be offset spatially in the detector relative to their direct image X and Y coordinates. For the UVIS array the offset varies as a function of the position due to geometric distortion \citep{Rothberg2011_UVISG280}. The relationship between the coordinates in x and y pixel position also needs to be taken into account when planning the observations in X and Y arcsecond coordinates (see WFC3 data handbook for conversion functions \footnote{\href{https://hst-docs.stsci.edu/display/WFC3IHB/C.2+WFC3+Patterns}{WFC3 Data Handbook Appendix C.2}}).

- \textbf{Orientation constraints:} The spectral trace of the positive and negative orders extend across over 500 pixels each depending on the brightness of the target. In a crowded field or where a target is part of a visual binary system, tight orient constraints need to be placed on the observations to prevent contamination from nearby stars. This is often mitigated in WFC3/IR grism observations using spatial scans where the spectra can be extracted by differencing the individual non-destructive reads within the final science frame. However, as WFC3/UVIS grism observations can only be conducted in stare mode, up-the-ramp sampling cannot be used to recover overlapping spectra. 

- \textbf{Cosmic rays:} The large wavelength coverage that extends significantly into the blue wavelengths increases the number of detected comic rays compared to the IR detectors.  

- \textbf{JWST challenges:} For JWST a number of the instrument modes that will be utilized for exoplanet timeseries data exhibit curved spectral traces, overlapping spectral orders, and contamination constraints from additional sources on the sky. NIRISS SOSS mode is most similar to the G280 grism with both strongly curved spectral traces and overlapping spectral orders. It is also expected that NIRSpec Bright Object Time Series observations will also have a slightly curved trace. For all slitless modes on JWST used for exoplanet time series observations contamination overlap will need to be carefully considered and orientation constrained carefully sampled.

% ------------------------
% Advantages
% ------------------------
\subsection{Advantages}
While we have detailed many challenges there are also significant advantages to this instrument over other modes in the NUV and optical. We detail these here. 

- \textbf{Wide wavelength coverage:} observations are conducted over the whole wavelength range 200--800\,nm in a single exposure. Low resolution spectra across this wide wavelength range can address the two main exoplanet science points revealed by HST observations; cloud opacities and atmospheric escape. The G280 grism can measure both the lower atmosphere sensitive to aerosol scattering, while large atmospheric escape signatures can be detectable in narrow bands around strong Fe and Mg signatures at $<$300\,nm. 

\begin{figure}
    \centering
    \includegraphics[width=\columnwidth]{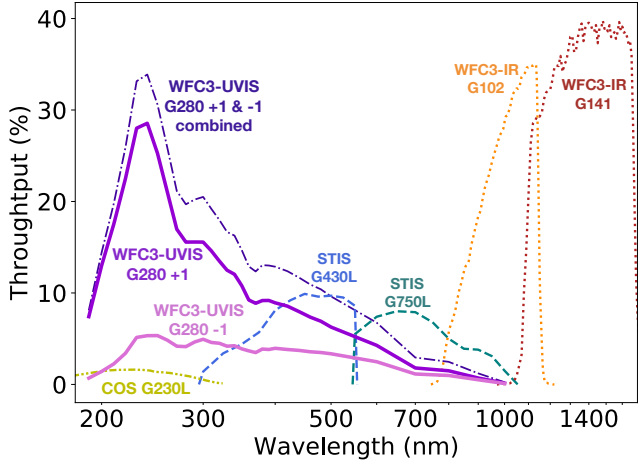}
    \caption{Throughput curves for HST instruments and modes commonly used for exoplanet time-series observations. Solid lines are the WFC3-UVIS G280 grism +1 and -1 orders; the dark dot-dashed line is the combined transmission of both orders. Dashed lines are STIS G430L and G750L gratings. -..- line shows the COS G230L. Dotted lines are WFC3-IR G102 and G141 grisms.}
    \label{fig:throughputs}
\end{figure}

- \textbf{Multiple spectral orders:} both the positive and negative orders are measured in each exposure. 
The UVIS CCD is split into two chips (1 \& 2), with each chip of 2051$\times$2048 pixels easily encompassing both spectral orders which each cover $\approx$500 pixels in the dispersion direction. 
In the presented observations we use chip 2 as it has been shown to be more stable than chip 1. We therefore also recommend the use of chip 2 for future studies. 

- \textbf{Throughput:} WFC3/UVIS has the highest throughput amongst all HST instruments in the wavelength range from its lower cut off at 200\,nm to the upper end at $\sim$800\,nm. The throughput of UVIS G280 in the NUV is on average 25 times that of STIS E230M between 200--400\,nm, and roughly four times that of STIS G430L at 350\,nm. UVIS G280 also has the advantage of being able to measure both positive and negative spectral orders that have a combined throughput greater than STIS G430L across the whole wavelength range (see Fig.\,\ref{fig:throughputs}).

- \textbf{New calibration program:} Prior to these observations there have been three instrument science reports \citep{Kuntschner2009_G280,Rothberg2011_UVISG280,Pirzkal2017_G280} and no scientific papers using this grism. As demand for time-series observations with this grism have increased there are now new calibration programs being implemented to better characterize the detector and improve the trace fitting for all spectral orders. Calibration of the instrument and mode are important to understand the structure of the CCD, on-sky changes in the PSF, and wavelength dispersion across the detector - especially under the requirements of long term stability for exoplanet investigations that span multiple HST orbits. \\

Overall the WFC3/UVIS G280 grism has many challenges that are difficult but not impossible to overcome, and a significant advantage over other instrument modes in this wavelength range. In the following sections we detail the observations taken and the measurements made with the tools to overcome these challenges.

% ------------------------
% SECTION:
% OBSERVATIONS
% ------------------------
\section{UVIS G280 Observations}\label{sec:obs}
We used HST's WFC3/UVIS channel with the G280 spectroscopic slitless grism to observe the spectrum of the transiting exoplanet host star HAT-P-41 from 200--800\,nm (GO-15288, PIs D.K. Sing \& N.K. Lewis). Unlike the WFC3/IR G102 and G141 grisms, the UVIS G280 grism produces a spectrum that is strongly curved, with overlapping spectral orders at longer wavelengths, and a dimmer ($\sim$60\%) -1 order spectrum compared to the +1 order. We designed an observation strategy that would cover both +1 and -1 orders simultaneously to examine this difference in flux and test the usability of the G280 grism for time-series exoplanet studies. 

We observed the target HAT-P-41, in the constellation of Altair, over two visits, each consisting of five HST orbits, to measure the transit of the hot Jupiter exoplanet HAT-P-41b. The two visits were separated by a single planetary orbital period (visit 1: 2018 August 1st; visit 2: 2018 August 4th, period\,=\,2.694047\,days), significantly reducing the potential impact of any stellar variations on the transits of this quiet F6 star.   

Each visit consists of 54 exposures, over 5 HST orbits, with exposure times of 190 seconds. We used a 2100\,$\times$\,800 sub-array, with a POS TARG Y offset of -50'' to center the spectrum on chip 2. The sub-array is cut out of the full 2051\,$\times$\,4096 pixel CCD which contains chip 1 and 2, where chip 1 and chip 2 are separated by 1.2''. 
Our target star, HAT-P-41, has a nearby companion separated by 3.615'', equivalent to $\approx$\,91.5 pixels on the detector. The nearby companion resulted in a number of tight orientation constraints on the observation. However, our sub-array is large enough to capture both full +1 and -1 spectral orders around the 0th order trace.  
The maximum flux obtained in a single pixel in the spectral trace is $\approx$\,36,000\,e$^-$, keeping it well within the saturation and non-linearity limit of the detector, which is approximately 67,000--72,000\,e$^-$ \citep{ISR2010-10}. 

\subsection{Spectral Extraction}\label{sec:extraction}
The spectral traces for both visits and both +1 and -1 orders were extracted using calibration files provided by the WFC3 team. 
A complete extraction and reduction of the provided data requires the following steps: 
\begin{enumerate*}[label={\alph*)},font={\color{red!50!black}\bfseries}]
\item cosmic ray removal 
\item background subtraction 
\item aperture determination, and 
\item trace fitting
\end{enumerate*}
We then use the WFC3 UVIS calibration files to compute the wavelength solution for each spectral order. We also performed spectral extraction with IRAF and custom IDL routines as a second check on the extraction procedure as this is the first published analysis of G280 grism data for time-series spectroscopy (see \ref{sec:IRAF} for details).

% \subsection{WFC3Tools Spectral Extraction}
% Tools for the calibration files can be found here in this tar file http://www.stsci.edu/~WFC3/grism-resources/WFC3.UVIS.G280.cal.tar.gz

% ------------------------
% Cosmic ray removal
% ------------------------
\paragraph{\textbf{Cosmic ray removal}}
We use the ``flt'' files from the \textit{Calwfc3} pipeline to analyze each exposure. Cosmic rays were then rejected by examining the time series for each pixel, and flagging and replacing 3.5-$\sigma$ outliers in an iterative process. We also applied a further spatial cosmic ray cleaning step by rejecting cosmic rays through Laplacian Edge Detection \citep{vandokkum2001}. We do a final cosmic ray cleaning on the extracted 1D stellar spectra by comparing them to the median combined spectra and replacing outliers which deviated by more than 3.5\,$\sigma$. Where cosmic rays are flagged temporally we replace the pixel value with a median of the time sampled pixel, where cosmic rays are spatially flagged a median of the surrounding pixels in the same frame is used.
\begin{figure}
    \centering
    \includegraphics[width=\columnwidth]{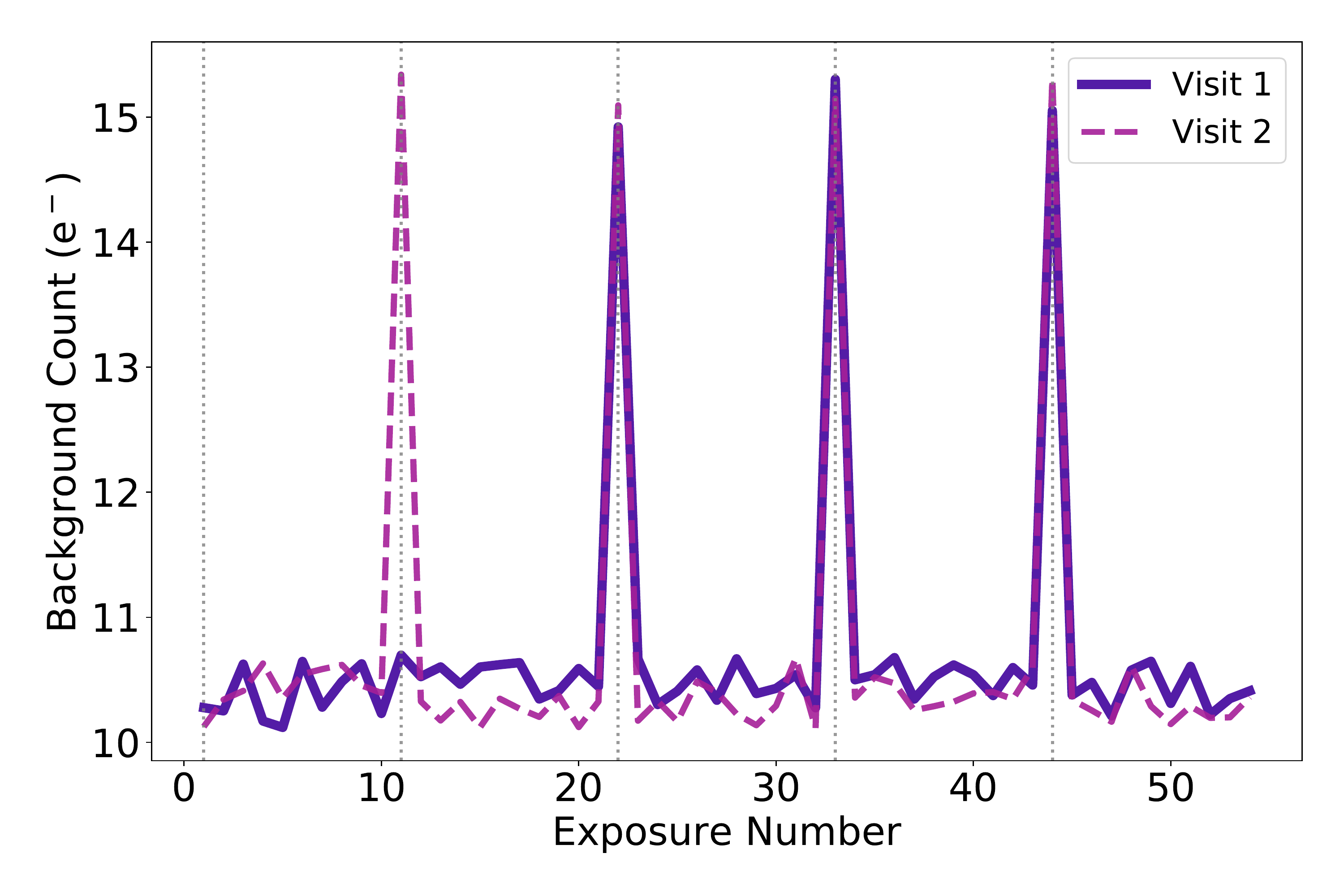}
    \caption{Modal background count for each exposure in visit 1 (solid) and visit 2 (dashed) across the whole sub-array. The dotted vertical lines indicate the start of a new HST orbit. The background is higher at the start of each HST orbit with a bi-model distribution, perhaps due to stray Earthshine or orbital effects on the telescope. }
    \label{fig:background}
\end{figure}

% ------------------------
% Background subtraction
% ------------------------
\paragraph{\textbf{Background subtraction}}
We use the local background estimation, similar to WISE (see section 4.4\,c \citealt{cutri2012_WISE} \footnote{\href{http://wise2.ipac.caltech.edu/docs/release/allsky/expsup/sec4_4c.html}{WISE All-sky release explanatory supplement, Section 4.4 c}}), by computing the pixel mode for each image and subtracting that from each pixel. The mode, or most common binned histogram value, tends to be robust against the bright tail of the distribution that is caused by other stars and cosmic-ray (or hot) pixels in the exposure. We compared this to the mean and median sigma clipped pixel values and found good agreement, giving weight to the mode being resistant to outliers. In each visit the first exposure of each orbit has much higher background than the other exposures with a slightly bi-modal distribution around the peak of the histogram (see Fig.\,\ref{fig:background}), perhaps due to stray Earthshine or orbital effects on the telescope. We remove the first exposure of each orbit in both visits in the lightcurve analysis. 

\begin{figure}
    \centering
    \includegraphics[width=\columnwidth]{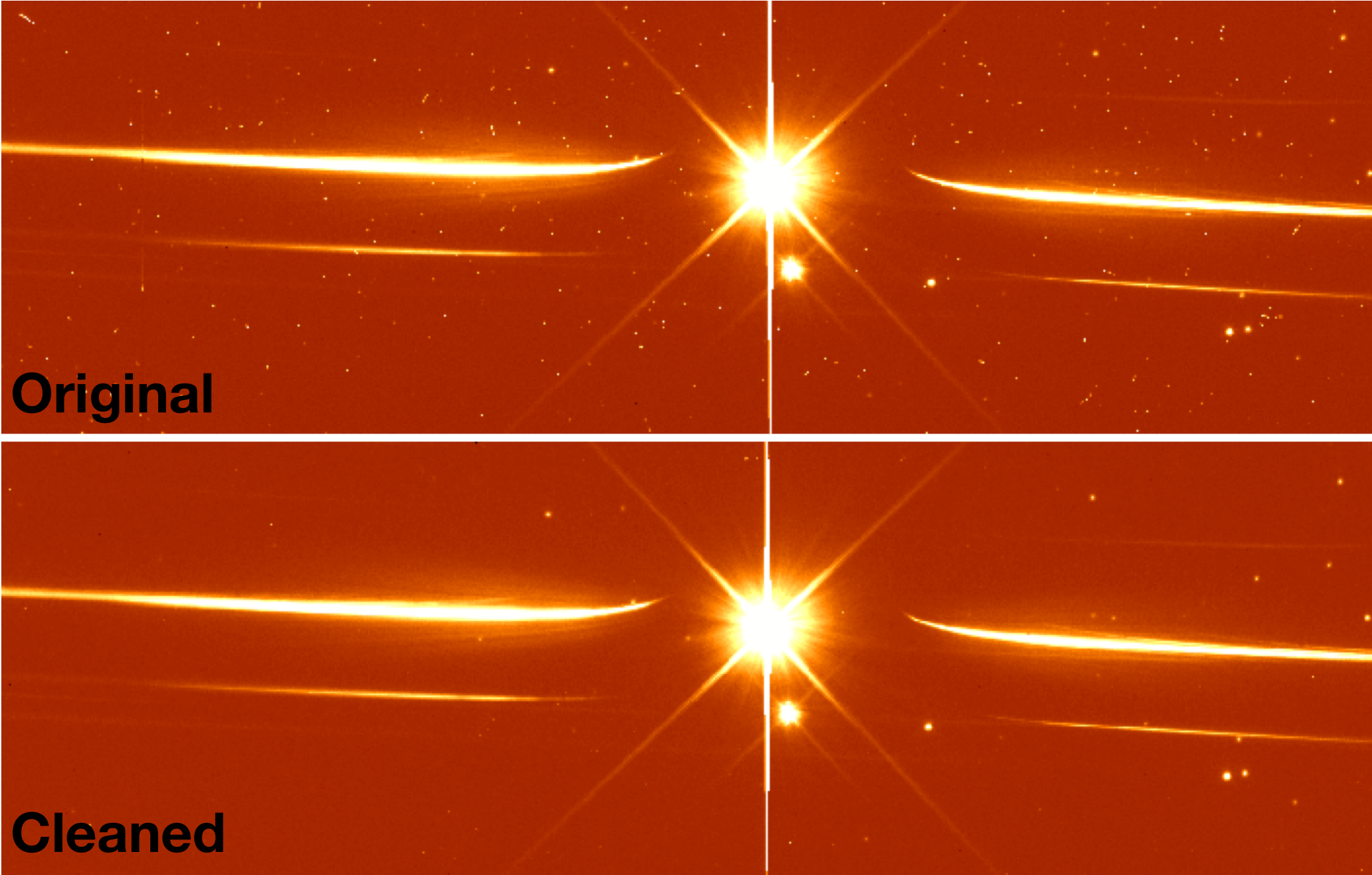}
    \caption{HST WFC3 UVIS/G280 spectral image. Top: ``flt'' file processed and available on the MAST archive. Bottom: cleaned file with cosmic rays and hot pixels removed, and flat fielding applied. In this comparison you can clearly see the difference between the original and cleaned data demonstrating the requirement for accurate and precise treatment of detector artifacts and cosmic ray hits.}
    \label{fig:cleaned_spec}
\end{figure}

Figure\,\ref{fig:cleaned_spec} shows the visual difference between the original ``flt'' images and a cleaned-background-subtracted exposure. We save the cleaned and background subtracted images as \textsc{fits} files to be used for the trace fitting routines.

% --------------
% Trace fitting
% --------------
\paragraph{\textbf{Trace fitting}}
To extract the target spectrum using the provided calibration files for UVIS G280\footnote{\href{http://www.stsci.edu/hst/instrumentation/wfc3/documentation/grism-resources}{G280 UVIS grism files}}, the sub-array image needs to be re-embedded into the full frame \citep{Rothberg2011_UVISG280}. This can be done using the \textsc{embedsub} routine in \textsc{wfc3tools}\footnote{\href{https://github.com/spacetelescope/wfc3tools}{https://github.com/spacetelescope/wfc3tools}}. This routine also requires the ``spt'' files be downloaded from the MAST database and contained within the same folder as the cleaned \textsc{fits} files generated from the previous steps. 
\begin{figure*}
    \centering
    \includegraphics[width=\textwidth]{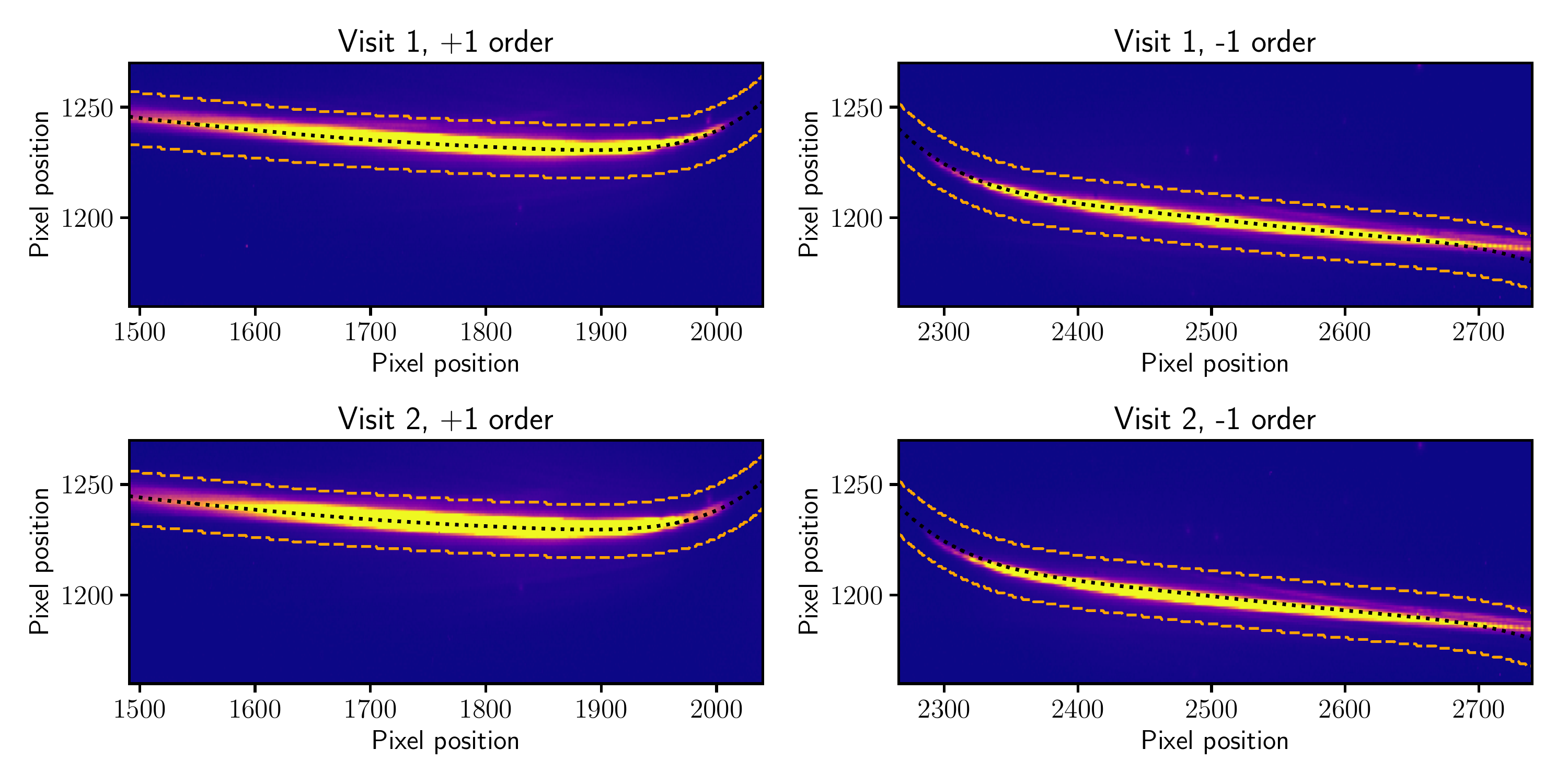}
    \caption{HST WFC3 UVIS/G280 spectral image. Top: visit 1 +1 spectral order (left) and -1 spectral order (right). Bottom: visit 2 +1 spectral order (left) and -1 spectral order (right). All images are background subtracted and cosmic rays have been removed. The dotted black line shows the calculated trace center, with the extent of the +-12 pixel aperture shown in orange dashed lines. At lower flux values the spectral trace does not fit quite as well but the full flux is captured inside the selected aperture. Color shows flux, truncated at 25\,e$^-$s$^{-1}$.}
    \label{fig:spec_trace}
\end{figure*}

Direct images of the target were taken with the F300X filter at the start of each visit to provide an accurate location of our target on the detector. Visits 1 and 2 were positioned on the detector within 1 pixel of each other with x, y centroid positions of [2040.8756, 1063.8825] and [2041.0399, 1062.9073] respectively. 

\begin{figure}
    \centering
    \includegraphics[width=\columnwidth]{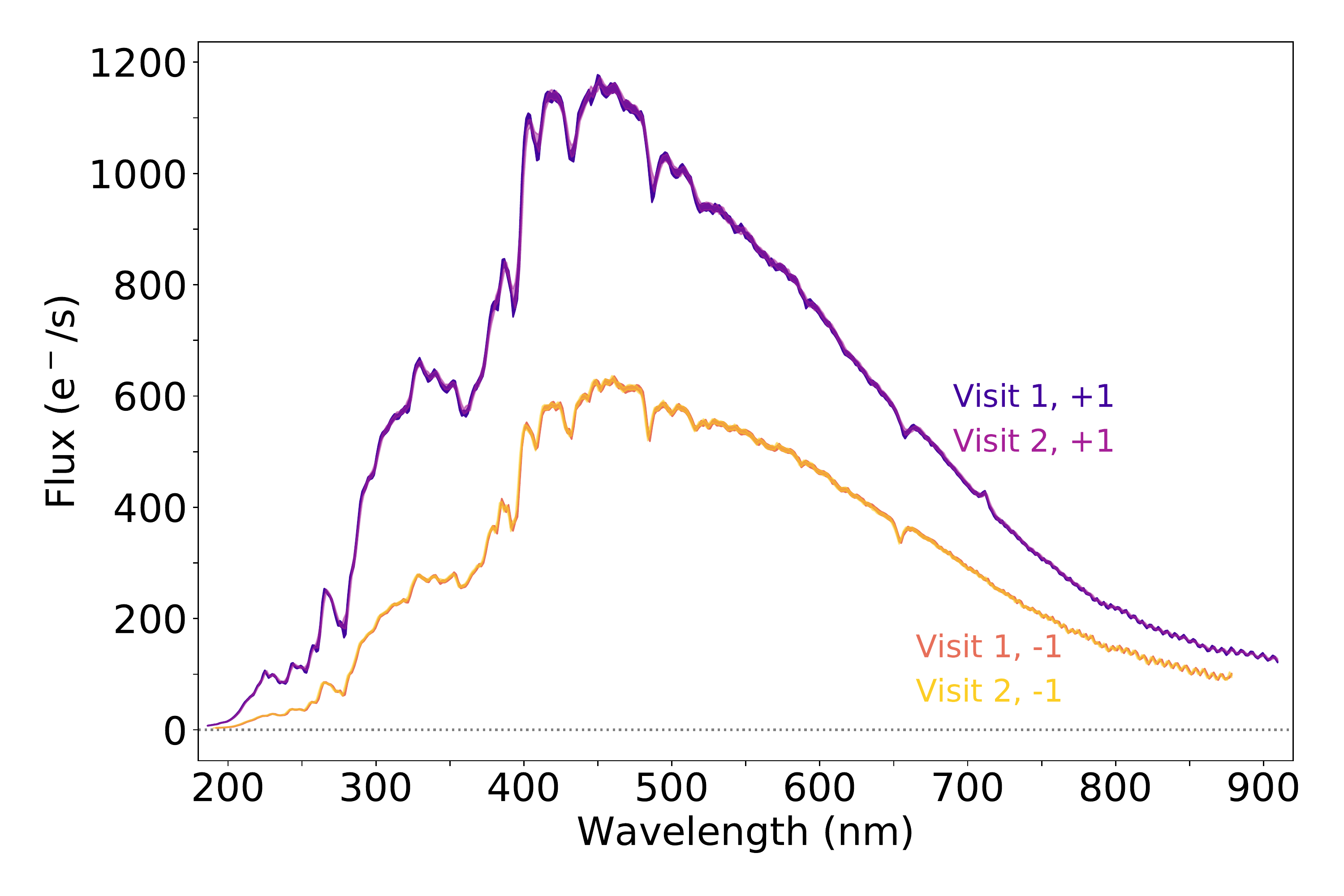}
    \caption{The 16--84 percentile range of each visit and orders spectral trace. The +1 orders and -1 orders overlap closely, making it difficult to tell the two visits apart and demonstrating the stability of the instrument and the star. The -1 orders are $\sim$50\% dimmer than the +1 orders, with little to no flux short of 250\,nm. Above 800\,nm fringing patterns can clearly be seen in the stellar spectra and we do not use these wavelengths for the lightcurve analysis.}
    \label{fig:stellar_spec}
\end{figure}

Using the description of the spectral trace of the G280 UVIS grism from \citet{Pirzkal2017_G280}, we computed the expected location of the trace in each exposure of our G280 datasets. In summary, \citet{Pirzkal2017_G280} compute the trace location as a function of the x-pixel on the detector and a high order 2D polynomial is fit across the trace. The best fit trace is defined by a 6th order polynomial function with a linear 2D field dependence.
The reference column for the polynomial fit is chosen to be close to the inflection point of the trace to ensure the best fit to both the highly curved spectrum at short wavelengths and the near-linear trace at longer wavelengths. The polynomial function reproduces the position of both the +1 and -1 spectral orders to within a fraction of a pixel from 200--800\,nm. Figure\,\ref{fig:spec_trace} shows the central trace position for both visits and computed for the +1 and -1 spectral orders. The trace fits are currently best calibrated to the +1 order, however, the authors note that there is a new WFC3/UVIS G280 calibration program that will fully characterize the -1 and additional spectral orders. At longer wavelengths, toward the tail end of the spectral trace, fringing effects come into play that divert the spectra from the fit polynomial trace (see Fig.\,\ref{fig:stellar_spec}).

A simple extraction of the spectrum contained in each dataset was created by adding up the observed count rates in pixels above and below the computed location of the trace. 
We tested apertures ranging from $\pm$5 pixels around the central trace to $\pm$50 pixels. To determine the best aperture we minimized the standard deviation of the residuals for out-of-transit counts. We find that the optimal aperture is $\pm$12 pixels (see Fig.\,\ref{fig:spec_trace}), to account for the slightly extended wings of the trace \citep{Kuntschner2009_G280}.
Both the +1 and -1 spectra orders were processed in this manner. 

The overlapping spectral orders are expected to impact the spectrum in the long wavelengths approximately beyond 400\,nm. However, these observations were not ideal to show the impact of overlapping spectral orders as the brightness of the star in the shorter wavelengths is too dim, $\approx$65$\times$ dimmer than the first order trace.  We discuss potential corrections to this in more detail in \S\,\ref{sec:spec_analysis}.

% ---------------------
% Wavelength solution
% ---------------------
\paragraph{\textbf{Wavelength solution}}
The wavelength solution is calculated from the trace position using the equation detailed in \citet{Pirzkal2017_G280} which is calibrated from 190 to 800\,nm. The extracted wavelength solution is good to +/- 0.7\,nm which is roughly half of a UVIS resolution element. We measure the mean spectral dispersion in the first order which varies from $\sim$1.1--1.6\,nm per pixel over the full spectral range 200--800\,nm. 

We plot the stellar spectra for both visits and first order spectra in Fig.\,\ref{fig:stellar_spec}, showing the 16-84 percentile range of each spectrum with remarkable agreement between visits, demonstrating the stability of the instrument. Beyond 800\,nm the target spectrum shows extreme fringing effects and is not calibrated, thus we remove it from this analysis. It is also clear to see that the -1 order is significantly dimmer across the whole wavelength range with a large impact on the short wavelengths, short of 250\,nm, where the flux drops to near-zero.  

% ------------------------
% Sub-section:
% IRAF spectral extraction
% ------------------------
\subsubsection{IRAF APALL Spectral Extraction}\label{sec:IRAF}
We also performed spectral extraction with IRAF and custom IDL routines. The images were first background subtracted and cosmic rays were removed in the same way as detailed above. We then used IRAF's APALL routine to extract the spectra for each image in the time series, finding an 8$^{th}$ order legendre polynomial was optimal for the spectral trace extraction as measured by the trace root mean square residuals. We note that with IRAF, the fixed aperture center varies smoothly to follow changes in the position of the spectrum across the dispersion axis and partial pixels are used at the ends. 
We extracted the spectra with a wide range of aperture sizes, finding a 24 pixel aperture was optimal. Similar to the UVIS calibration pipeline routines, the extracted spectra still exhibited a few cosmic rays not cleaned in previous processes, we also then perform the 1D stellar spectra cosmic ray removal step. Using IRAF APALL we were unable to replicate the wavelength solution calculation and therefore used the one calculated following \citet{Pirzkal2017_G280} that required the trace fitting following the UVIS calibration pipeline.\\

Both spectral extraction techniques produce near identical stellar spectra and transmission spectra. However, in the following sections we adopt and present the analysis based on the spectra extracted using the UVIS calibration pipeline as it is widely accessible,  publicly available extraction method that does not rely on proprietary custom routines, and has a fully consistent wavelength solution. 

% -----------------------
% SECTION:
% BROADBAND ANALYSIS
% -----------------------
\section{Broadband white-light analysis}\label{sec:analysis}
Prior to measuring the transmission spectrum of HAT-P-41b, we first analyze the broadband white lightcurve from 200--800\,nm.
In this section we detail the analysis of the broadband whitelight transit depth measured in the UVIS G280 transits for each visit and spectral order based on two different systematic treatment methods - instrument systematic marginalization \citep{wakeford2016} and jitter decorrelation \citep{sing2019}. 

% ------------------------
% Instrument systematic marginalization
% ------------------------
\paragraph{\textbf{Instrument systematic marginalization}} uses a pseudo-stochastic grid of corrective systematic models to measure the desired lightcurve parameters, namely the transit depth, via an evidence-based weight assigned by the data to each potential systematic model. 
We run a grid of 50 systematic models in the form of an extended polynomial;
\begin{equation}
    S(\mathbf{x}) = t_1\phi_t \times \sum^n_{i=1} p_i \phi^i_{HST} \times \sum^n_{j=1}l_j \delta^j_\lambda + 1
\end{equation}
where $\phi_t$ is the planetary phase representing a linear slope over the whole visit, $\phi_{HST}$ is the HST orbital phase accounting for ``HST thermal breathing'' effects, and $\delta_\lambda$ is the positional shift in the wavelength direction on the detector over the visit. Each of these parameters have scaling factors with the linear slope defined by $t_1$, and ``HST breathing'' and positional shifts fit up to a 4th order polynomial function defined by $p_{1-n}$ and $l_{1-n}$, respectively. Each of the scaling parameters are then either fit as free parameters to activate the systematic model or fixed to zero. The whole grid of 50 systematic models used in this analysis can be found in Table 2 of \citet{wakeford2016} note the table is 0 indexed.

We approximate the evidence (marginal likelihood) of each systematic model fit to the data using the Akaike Information Criterion (AIC). We then calculate the evidence-based weight ($W_q$) across all 50 systematic models and use the information from all models to marginalize over the desired parameter ($\alpha_q$). 
\begin{equation}
    \alpha_m = \sum^{N_q}_{q=0} (W_q \times \alpha_q)
\end{equation}
Equation (15) of \citet{wakeford2016}, where $N_q$ is the number of models fit, and $\alpha_m$ is the resulting marginalized parameter. The uncertainty is then calculated in a similar way based on the weights (see Equation (16) of \citealt{wakeford2016}).

\begin{figure}
    \centering
    \includegraphics[width=\columnwidth]{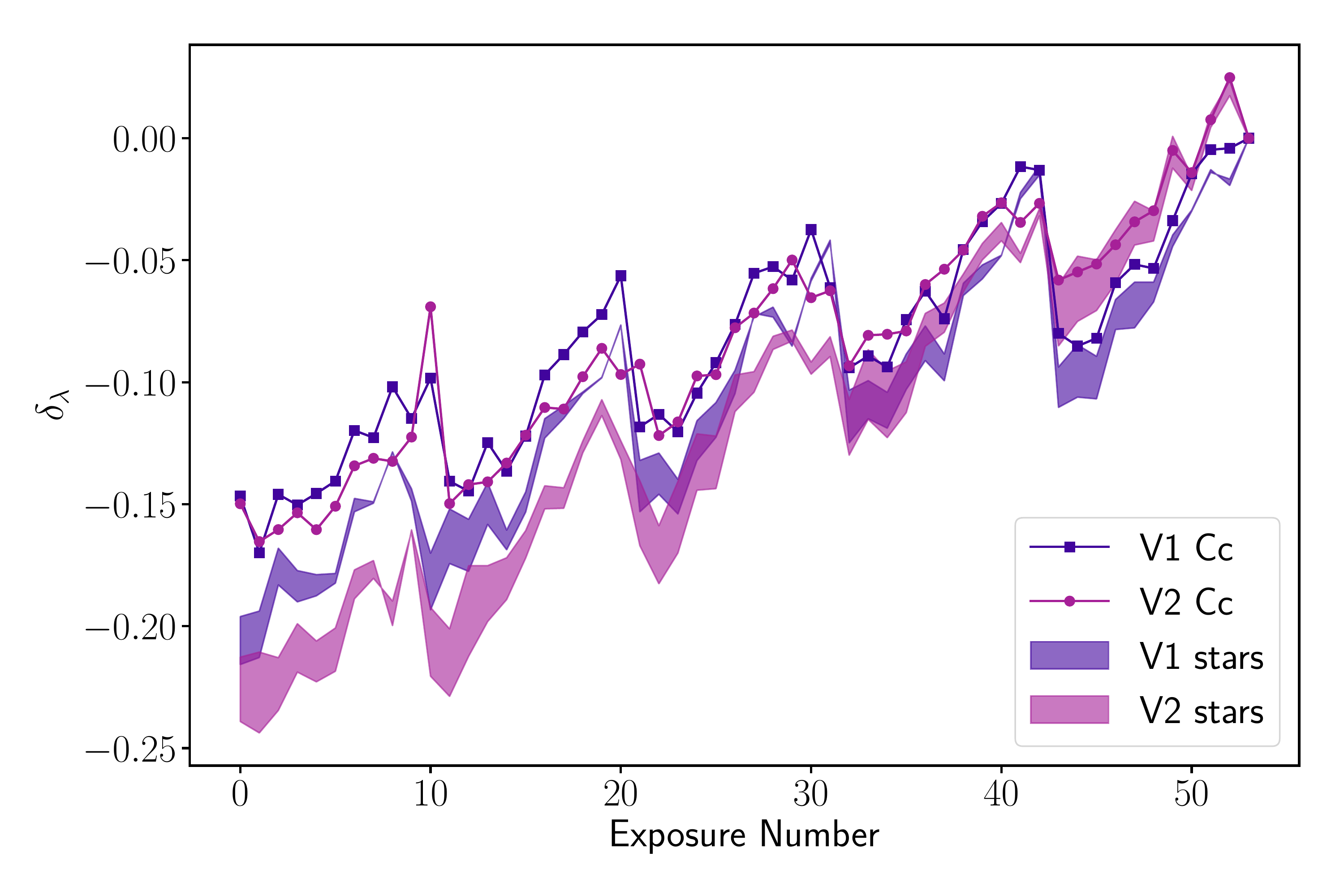}
    \caption{Spectral position changes over the course of each visit, measured by cross-correlating to a template spectrum (points, Cc), and by fitting background sources on the full exposure image (shaded regions, stars). Each are shown relative to the final exposure for comparison. The spectral shifts are accounted for in the systematic treatment of each lightcurve.}
    \label{fig:sh}
\end{figure}

\begin{figure}
    \centering
    \includegraphics[width=\columnwidth]{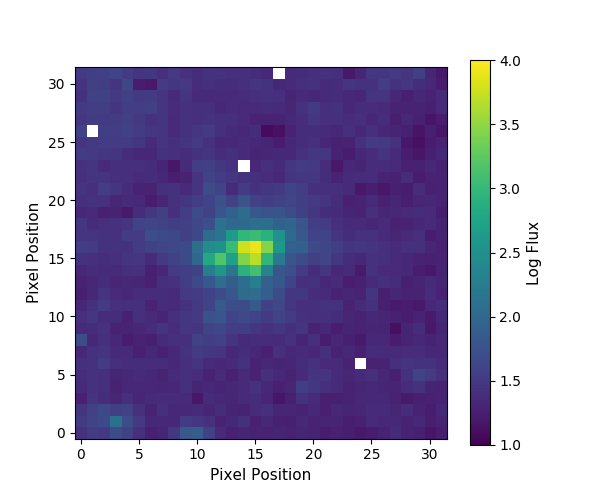}
    \caption{Mean flux in \textit{Spitzer}'s 3.6\,$\mu$m channel. Plotting on a logarithmic scale reveals HAT-P-41's faint, nearby companion at pixel position (12,15). We limit our photometry aperture size to 2.25 pixels to minimize contamination from the companion. Bad pixels are masked in white.}
    \label{fig:spitzer_mean}
\end{figure}
% ------------------------
% Jitter decorrelation
% ------------------------
\paragraph{\textbf{Jitter decorrelation}} uses HST's Pointing Control System to detrend photometric time-series data. Based on the results of \citep{sing2019}, we include optical state vectors traditionally used for STIS \citep{sing2011b} as well as several jitter vectors.  The full systematics model, $S(\mathbf{x})$, used to detrend the lightcurve is written as,
\begin{multline} 
S(\mathbf{x}) = p_1\phi_{t} + \sum^4_{i=1}p_{i+1}\phi^i_{HST} + p_6\delta_\lambda \\ 
% p_2\phi_{HST}+p_3\phi_{HST}^2+p_4\phi_{HST}^3+p_5\phi_{HST}^4+\\
% p_6\delta_{\lambda}
+ p_7X_{psf} + p_8Y_{psf} + p_9RA + p_{10}DEC\\
+ p_{11}Vn_{roll} + p_{12}Vt_{roll} + 1,
\label{eq:sys}
\end{multline} 

\noindent where $\phi_{t}$ is a linear baseline time trend, 
$\phi_{HST}$ is the 96 minute {\it HST} orbital phase,
$X_{psf}$ and $Y_{psf}$ are the detector
positions of the PSF as measured by the spectral trace,
$\delta_{\lambda}$ is the wavelength shift of each spectra as measured by cross-correlation, $V2\_roll$ and $V3\_roll$ are roll of the telescope along the V2 and V3 axis, 
$RA$ and $DEC$ are the right ascension and declination of the aperture reference, 
and $p_{1..12}$ are the fit systematic parameter coefficients. The first portion of this function was found to be the best functional form of the additional systematic features and corresponds to one of the models used in the marginalization grid. This function is then fit for all transit lightcurves in this form and is not marginalized over to determine the optimal functional form in each lightcurve. 
The full jitter decorrelation set results in up to twelve total terms used to describe the instrument systematics of the dataset in question.  However, in practice not all of these parameters are needed. For each visit and each of the two orders, we used the AIC and measured red noise, $\sigma_{\rm r}$, to determine the optimal optical state vectors to include from the full set without over-fitting the data and minimizing the red noise. \\

Both systematic marginalization and jitter decorrelation require a measurement of the spectral positional changes on the detector across the duration of the observation ($\delta_\lambda$). To calculate the shift, we cross-correlate the 1D stellar spectra to a template spectrum and measure the displacement across the whole wavelength range. To demonstrate that this accurately represents the physical shift on the detector, we measured the position for three background sources distributed across the exposure image. We selected the most Gaussian-like sources from the full image and used a 2D-Gaussian fit to their 0th order spectrum in each exposure of each visit. In this case we cannot use the 0th order of the target or its stellar companion to measure this shift as they are both saturated on the detector. Figure\,\ref{fig:sh} shows $\delta_\lambda$ for visits 1 and 2 measured using the cross-correlation method (Cc) and the range of positional values measured from the three background sources (stars). The form of the positional shifts are very similar with the vertical breaks showing where the telescope is reset after each HST orbit. The magnitude of the positional shifts is on the sub-pixel scale and is easily accounted for with either of the  systematic treatments detailed. Using the 2D-Gaussian fit to the background sources, we find that positional shifts in the y-direction are negligible and do not improve the fit to the data. 

Due to the phase coverage of HST observations, resulting from Earth occultation events, we are unable to accurately fit for the inclination, a/R$_*$, and orbital period of the system. 
Unfortunately, HAT-P-41b was not observed by the Transiting Exoplanet Survey Satellite (TESS) which would have allowed us to easily constrain the system parameters. To fit for these vital parameters we instead use two transit observations from the Spitzer Space Telescope IRAC instrument to obtain accurate system parameters for the inclination and a/R$_*$ of HAT-P-41b, detailed in \S\ref{sec:spitzer}. 
In \S\ref{sec:o-c} we present the measured center of transit times for these and previous transit observations of HAT-P-41b to determine the period of the planet, and in \S\ref{sec:wl_results} we present the results of the UVIS G280 broadband lightcurves for the two visits and for each spectroscopic order using both systematic treatments.

% ------------------------
% Sub-section:
% SPITZER
% ------------------------
\begin{figure*}
    \centering
    \includegraphics[width=\textwidth]{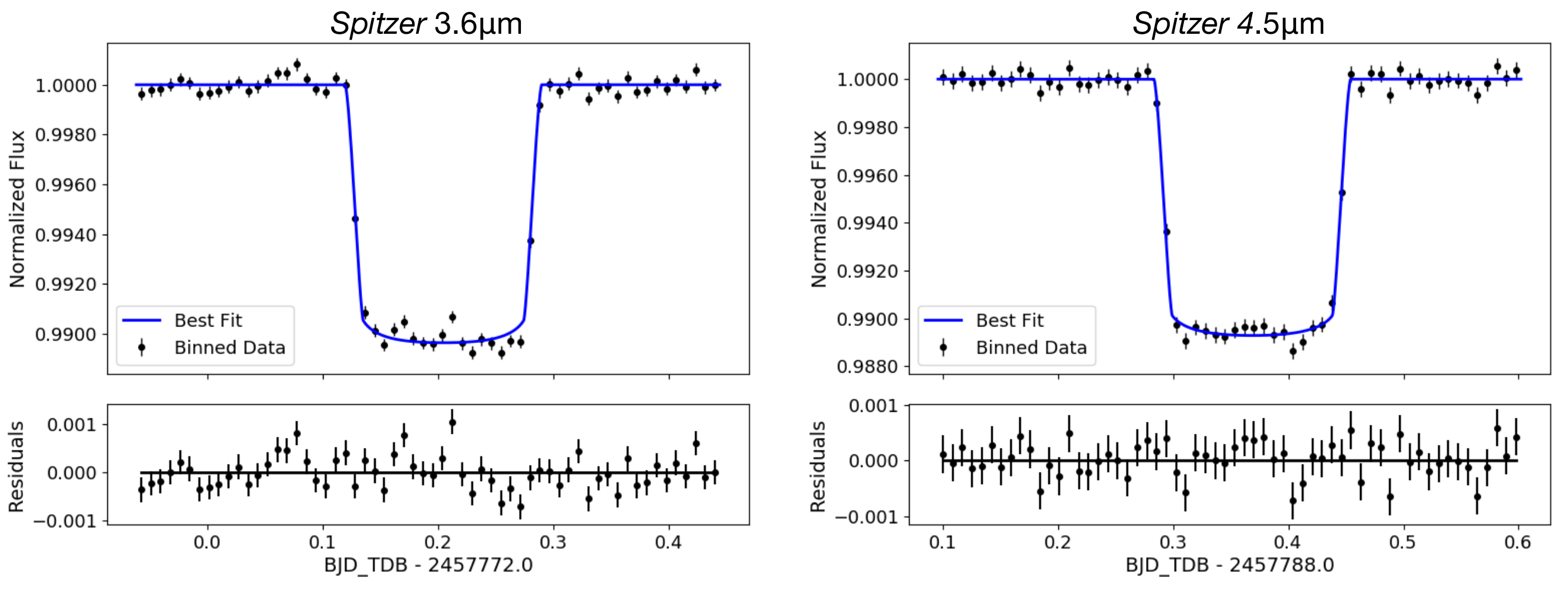}
    \caption{Transit light curves of HAT-P-41b using \textit{Spitzer}'s 3.6\,$\mu$m (left) and 4.5\,$\mu$m (right) channels.  We bin the data for plotting purposes only.  The 3.6\,$\mu$m residuals demonstrate a small amount of correlated noise at timescales shorter than the transit duration.
    }
    \label{fig:spitzer_lightcurve}
\end{figure*}

\subsection{Spitzer Data Analysis}\label{sec:spitzer}

\textit{Spitzer} program 13044 (PI: Deming) acquired transit observations of HAT-P-41b at 3.6 and 4.5\,$\mu$m on 2017 January 18 and 2017 February 3, respectively. The IRAC instrument \citep{IRAC} acquired 32$\times$32 pixel subarray frames at 2 second intervals in batches of 64. Each observation acquired a total of 21,632 frames over a span of $\sim$12 hours.

Using the POET pipeline \citep{Stevenson2011, Cubillos2013}, we apply a double-iteration, $4\sigma$ outlier rejection routine, 2D Gaussian centroiding, and $5\times$ interpolated aperture photometry over a range of aperture sizes.  
We convert times to BJD$_\mathrm{TDB}$ using the JPL Horizons interface.

We find that the best aperture size (as defined by the lowest standard deviation of the normalized residuals) is 3.0 pixels; however, at this size there is noticeable contamination from the nearby binary companion. This is evidenced by the correlation between aperture size and transit depth (significant at $3.3\sigma$).  HAT-P-41's stellar companion is located $\sim 3$ pixels away, in the wings of the primary star's point response function. This is shown in Figure \ref{fig:spitzer_mean}, where we depict the mean flux at 3.6\,$\mu$m on a logarithmic scale. We find that the impact of the stellar companion on the measured transit depth is minimal ($<1\sigma$) for apertures $\leq 2.25$ pixels and, thus, adopt this value for our final analyses. We note that the transit time, inclination, and semi-major axis parameters do not vary with our choice of aperture size.  

To derive our best-fit values (see Tables \ref{tab:system_params} and \ref{tab:ephemiris}), we fit both \textit{Spitzer} channels simultaneously using the transit model described by \citet{mandelagol2002}, a linear trend in time, and a BLISS map \citep{Stevenson2011} to account for intrapixel sensitivity variations.  We estimate uncertainties using the Differential-Evolution Markov Chain Monte Carlo technique \citep{terBraak2008} and test for convergence using the Gelmin-Rubin statistic \citep{Gelman1992} by ensuring that the potential scale reduction factor is within 1\% of unity. Figure \ref{fig:spitzer_lightcurve} shows \textit{Spitzer}'s normalized light curves and residuals.  The best-fit 3.6 and 4.5~{\micron} transit depths are $0.992\,{\pm}\,0.008$\,\% and $1.028\,{\pm}\,0.013$\,\%, respectively.

%After running a joint fit with both channels, I get the following median parameters and 1-sigma uncertainties:
%inc    = 89.2 +/- 0.6 degrees
%a/Rs   = 5.55 +/- 0.04
%CH1 Depth  = 0.992 +/- 0.008%
%CH2 Depth  = 1.028 +/- 0.013%

\begin{figure}
    \centering
    \includegraphics[width=\columnwidth]{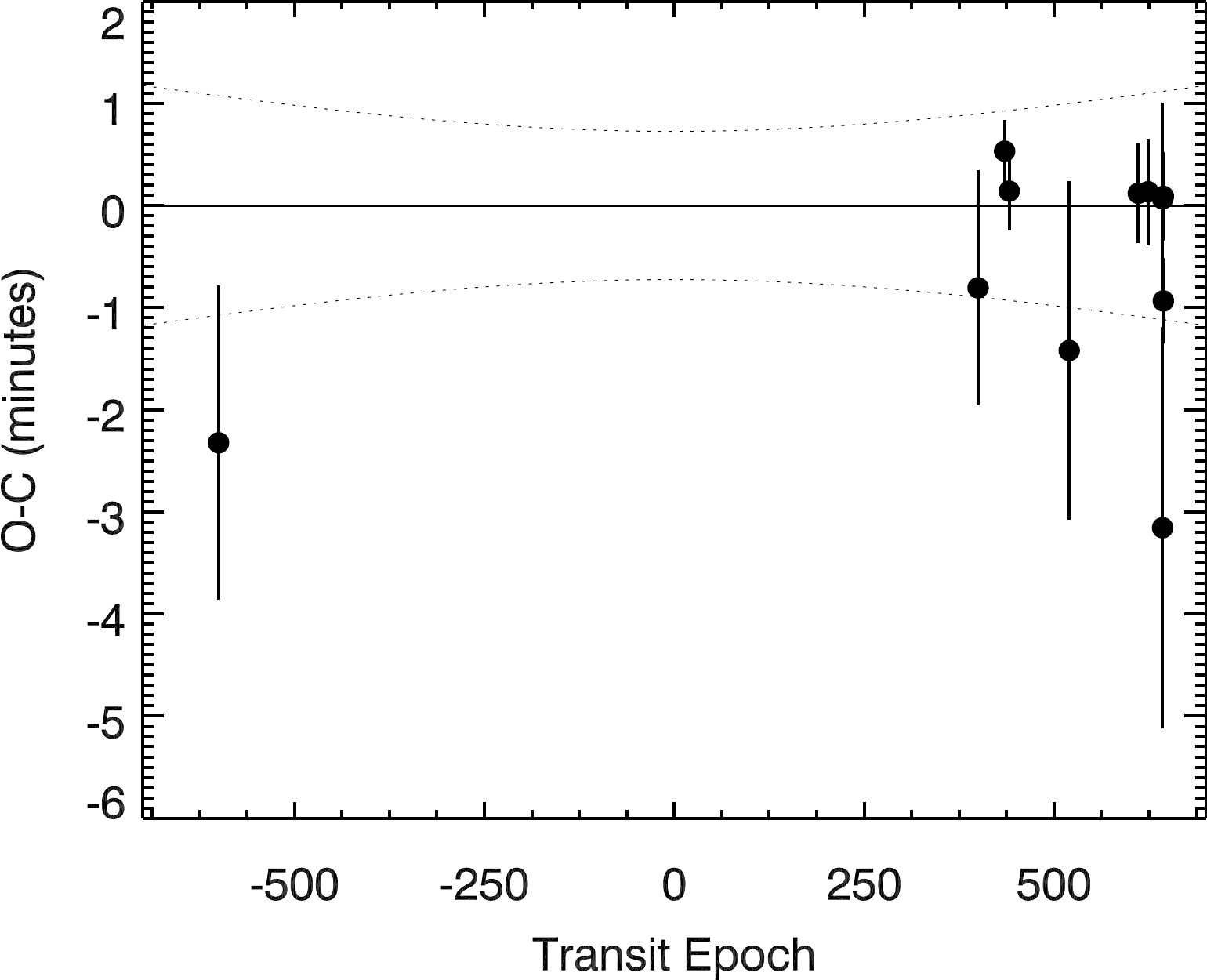}
    \caption{Observed minus calculated (O-C) diagram of measured HAT-P-41b transit times. The dashed line shows the 1-sigma uncertainty.}
    \label{fig:timing}
\end{figure}

\begin{table}
    \centering
    {\footnotesize
        \caption{Star and Planet parameters used in the lightcurve fitting process for this analysis.}
        \begin{tabular}{lll}
        \hline
        Parameter & Value & Reference \\ 
        \hline 
        \hline
        \textbf{Star}\\
        \hline
        V (mag) & 11.087 & \citet{hartman2012} \\
        M$_*$ (M$_\odot$) & 1.418 & \citet{hartman2012} \\
        R$_*$ (R$_\odot$) & 1.786 & \citet{morrell2019} \\
        Teff (K) & 6340 & \citet{morrell2019} \\
        {[Fe/H]} (dex) & 0.21 & \citet{hartman2012} \\
        log(g) & 4.14 &  \citet{hartman2012} \\
        \hline
        \textbf{Planet}\\
        \hline
        M$_p$ (M$_J$) & 0.795 & \citet{bonomo2017} \\
        R$_p$ (R$_J$) & 1.685 & \citet{hartman2012} \\
        Period (days) & 2.69404861 $\pm$0.00000092 & This work \\
        T$_0$ (days) & 2456600.29325$\pm$0.00050 & This work \\
        inclination ($^\circ$) & 89.17 $\pm$ 0.62 & This work \\
        a/R$_*$ & 5.55 $\pm$ 0.04 & This work \\
        ecc & 0.0 & \citet{bonomo2017} \\
        \hline
        \end{tabular}
    }
    \label{tab:system_params}
\end{table}

% ------------------------
% SUB-SECTION
% O-C
% ------------------------
\subsection{Updated Orbital Ephemeris}\label{sec:o-c}
We used previous and current data to calculate an up-to-date orbital period for HAT-P-41b, including the ephemeris from the discovery \citep{hartman2012}, as well as HST and \textit{Spitzer} transit data (see Table \ref{tab:ephemiris}).  The HST data includes the WFC3/UVIS transits where the +1 and -1 orders were treated independently (see \S\ref{sec:wl_results}), as well as WFC3/IR and STIS transits from the Hubble PanCET program (GO-14767, PIs D.K. Sing \& M. Lopez-Moralez, \citealt{sheppard2020}).
We converted all of the available transit times to BJD$_{\rm TDB}$ using the tools from \citep{2010PASP..122..935E}. These times were fit with a linear function of the period $P$ and transit epoch $E$,
\begin{equation}
T(E) = T_0 + EP.
\end{equation}
The resulting ephemeris is given in Table \ref{tab:ephemiris}, with the linear function giving a reasonable fit to the data (see Fig. \ref{fig:timing}), with a $\chi^2$ value of 14.47 for 9 degrees of freedom (DOF).

\begin{table*}[]
    \centering
    {\footnotesize
        \caption{Center of transit times used in Fig.\ref{fig:timing} to calculate the period of the planetary orbit as well as the resulting best-fit orbital ephemeris. All times have been converted to BJD$_{TBD}$.}
        \begin{tabular}{llll}
        \hline
        Instrument & Mode & Epoch & Note \\ 
        & &(BJD$_{\rm TDB}$) (days)&\\
        \hline 
        \hline
            &        & 2454983.86247  $\pm$ 0.00107  & \citet{hartman2012}  \\
HST WFC3-IR  & G141  & 2457677.912139 $\pm$ 0.0008   & \\%;WFC3/G141 (BJD_TDB)
Spitzer IRAC & CH1   & 2457772.20477  $\pm$ 0.00021  & \\ %Spitzer (BJD_TDB) - 7.9861E-4
Spitzer IRAC & CH2   & 2457788.36879  $\pm$ 0.00027  & \\%Spitzer (BJD_TDB)
HST STIS & G430L     & 2458001.197547 $\pm$ 0.001151 & visit 1  \\ %G430L Visit 83
HST STIS & G430L     & 2458246.357040 $\pm$ 0.000339 & visit 2  \\ %G430L
HST STIS & G750L     & 2458281.379682 $\pm$ 0.000363 &  \\ %G750L
HST WFC3-UVIS & G280 & 2458332.566558 $\pm$ 0.000656 & Visit 1, +1 order  \\ %UVISA1 visit 1
HST WFC3-UVIS & G280 & 2458332.564321 $\pm$ 0.001366 & Visit 1, -1 order \\%UVISB01 visit 1 -1
HST WFC3-UVIS & G280 & 2458335.260623 $\pm$ 0.000303 & Visit 2, +1 order \\%UVISA2 Visit 2
HST WFC3-UVIS & G280 & 2458335.259912 $\pm$ 0.000290 & Visit 2, -1 order \\%UVISB2 Visit 4
        \hline
Period $P$ (days) & & $T_0$ (BJD$_{\rm TDB}$) (days)  \\
2.69404861$\pm$0.000000918 & & 2456600.293253 $\pm$ 0.000504\\
\hline
        \end{tabular}
    }
    \label{tab:ephemiris}
\end{table*}

% ------------------------
% SUB-SECTION
% RESULTS
% ------------------------
\subsection{UVIS G280 Broadband Lightcurve Results}\label{sec:wl_results}
\begin{figure}
    \centering
    \includegraphics[width=\columnwidth]{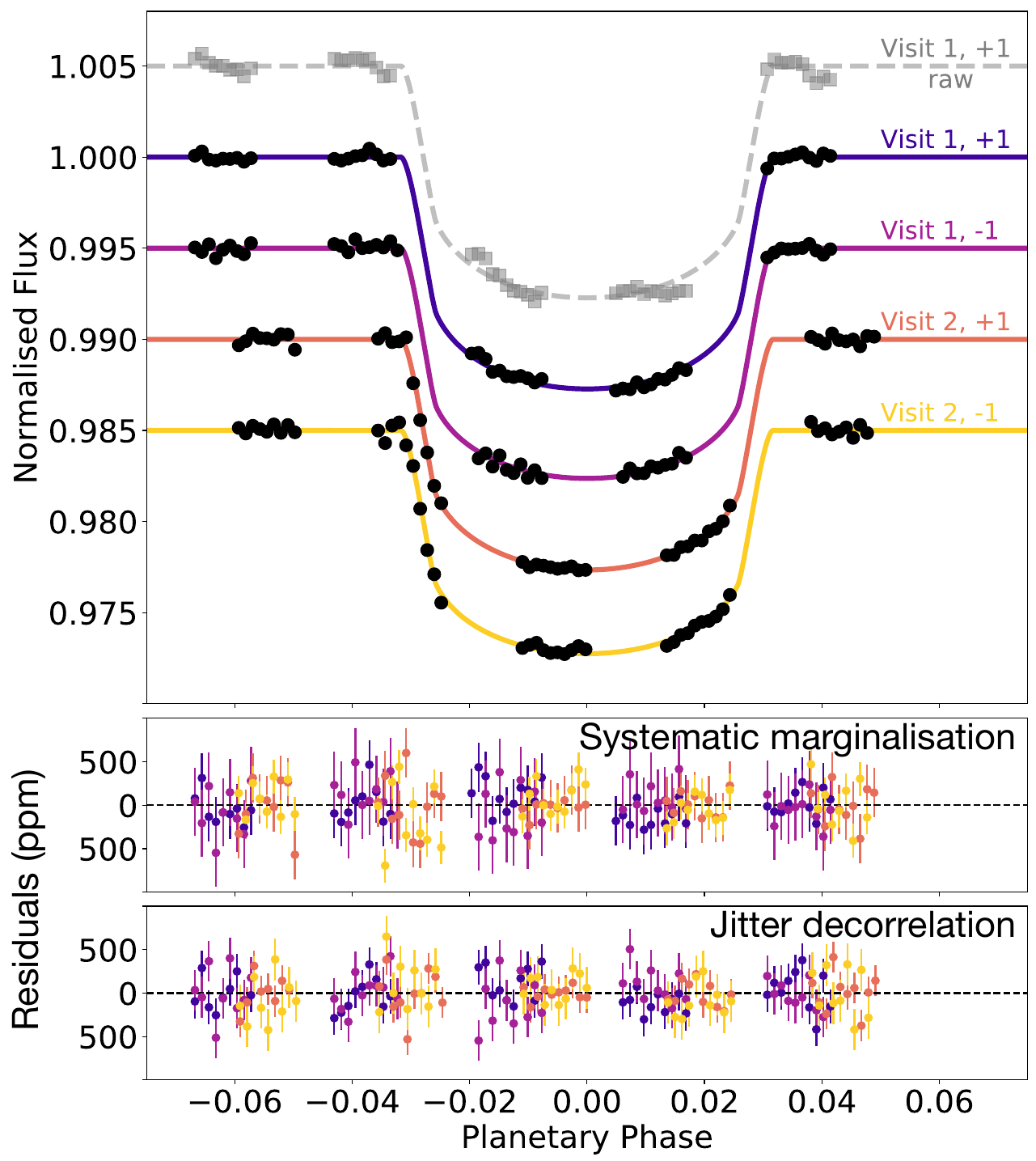}
    \caption{Top: broadband lightcurves. We show the raw extracted lightcurve for the visit 1 +1 spectral order to demonstrate the stability of the 1st HST orbit in the time series (light grey). The systematic corrected and normalized white lightcurves for each visit and spectroscopic order (colored labeled points) with the best fit transit model. Each point represents a single exposure. Each lightcurve is offset for clarity. Middle: residuals from each ligthcurve fit using the systematic marginalization method. Bottom: residuals for each lightcurve fit using the jitter decorrelation method. We measure the combined transit depth of HAT-P-41b to be $(R_p/R_*)^2$ = 1.0406\,$\pm$\,0.0029\,\% (SDNR = 221\,ppm) and 1.0330\,$\pm$\,0.0033\,\% (SNDR = 281\,ppm), for each method respectively.}
    \label{fig:whitelight}
\end{figure}

\begin{figure}
    \centering
    \includegraphics[width=\columnwidth]{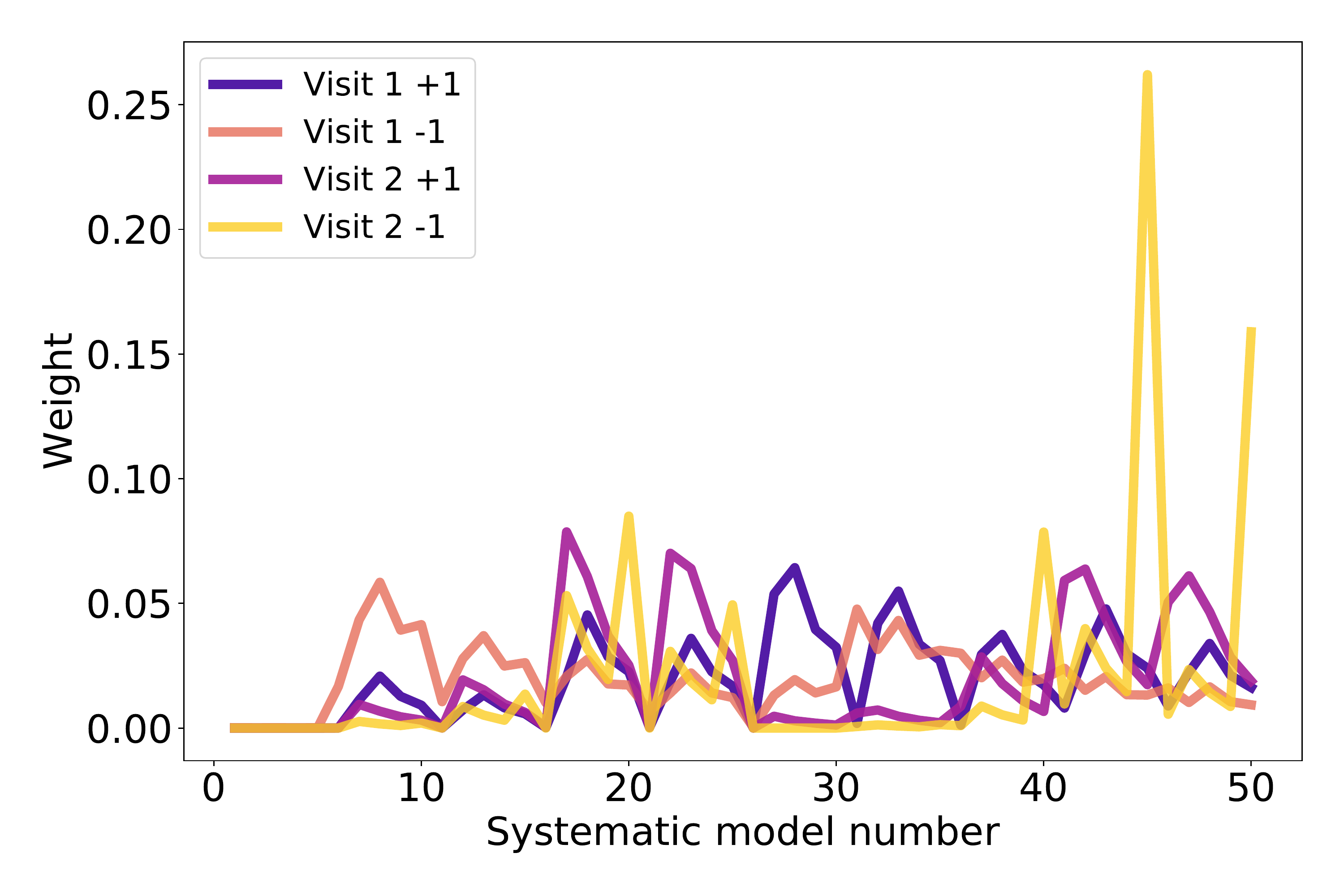}
    \caption{The evidence-based weight for each systematic model used in instrument systematic marginalization for each visit and order for the broadband lightcurve analysis. The table of systematic models relating to each number can be found in \citet{wakeford2016}.}
    \label{fig:weight}
\end{figure}
We measure the broadband transit depth for UVIS G280 by summing the flux from 200--800\,nm and correcting for systematics via systematic marginalization and jitter decorrelation independently for both visits and both spectral orders. We measure a combined transit depth of all four transit timeseries measurements of $(R_p/R_*)^2$~=~1.0406\,$\pm$\,0.0029\,\% and 1.0330\,$\pm$\,0.0033\,\%, with an average standard deviation on the residuals of 221\,ppm and 281\,ppm, using the systematic marginalization and jitter decorrelation methods respectively. There is a 1.7$\sigma$ difference between the two methods, likely due to the small differences between the uncertainties on each exposure for each analysis method that can be seen by comparing the bottom two panels of Fig.\,\ref{fig:whitelight}. 
In each analysis we use the same extracted stellar spectra, the same limb-darkening coefficients derived using the 3D stellar models presented in \citet{Magic2015}, and the same system parameters shown in Table\,\ref{tab:system_params}.

We show the four transit lightcurves (2 visits + 2 orders) corrected in Fig.\,\ref{fig:whitelight}. The lightcurves shown have been corrected using the most favored model applied in systematic marginalization, with the underlying models derived from the same most-likely systematic model. For both data analysis methods, systematic marginalization and jitter decorrelation, the transit model is fit iteratively with the systematic model to measure the transit depth. We note that the lightcurves in Fig.\ref{fig:whitelight} only represents a portion of the information obtained through marginalization as all the information from corrected data using other weighted systematic models also go into the final marginalized transit depth measurement (contribution weights can be seen in Fig. \ref{fig:weight}). Using jitter decorrelation, we derive a single solution for the lightcurve corrections and transit depth for each visit and spectral order. The individual lightcurves from jitter decorrelation are indistinguishable by eye compared to the systematic marginalization ones presented here. For a more direct comparison we show the residuals of both systematic analyses at the bottom of Fig.\,\ref{fig:whitelight} with their related uncertainties, both achieving near photon noise precision.

\begin{figure*}
    \centering
    \includegraphics[width=\textwidth]{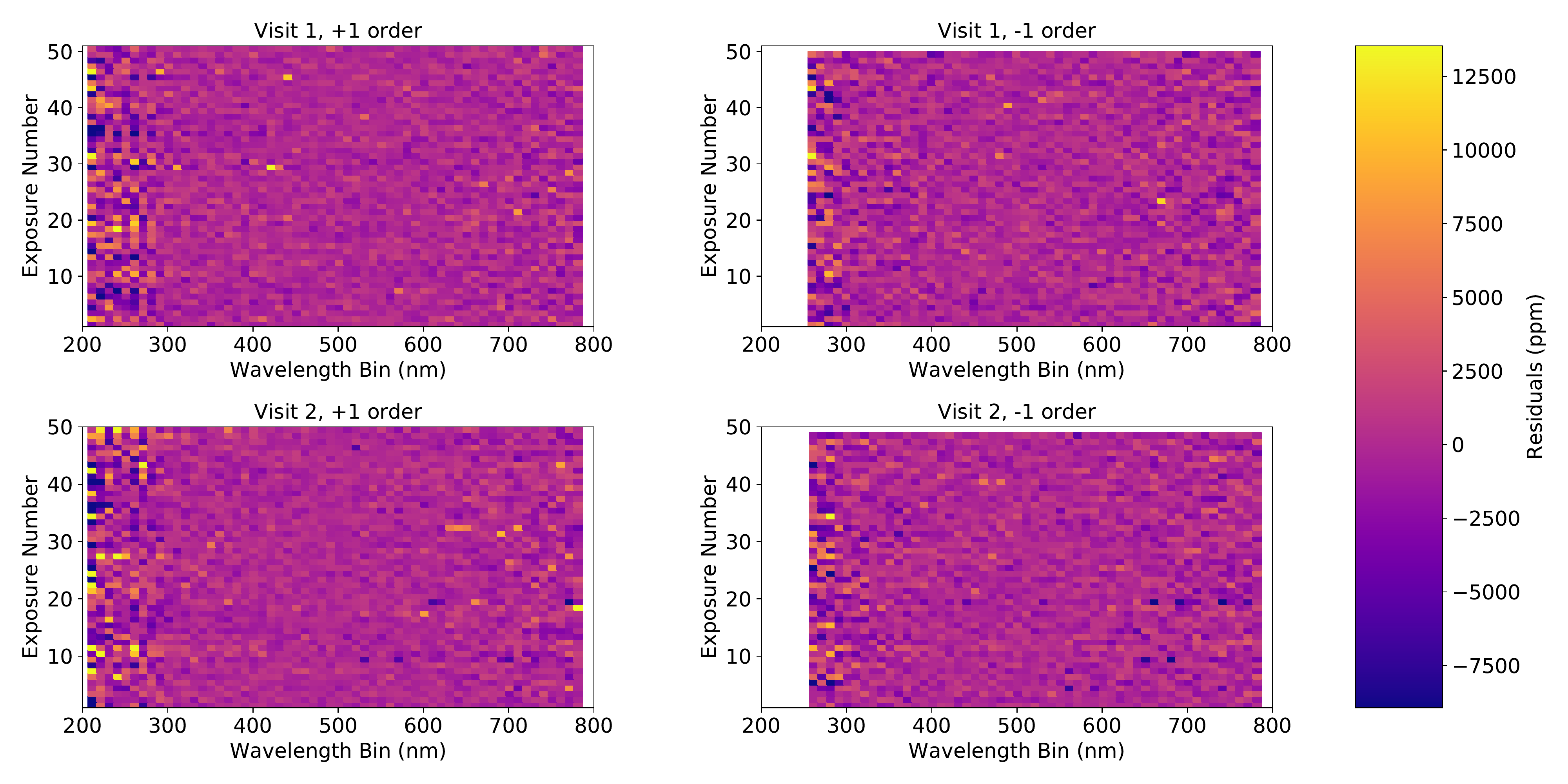}
    \caption{Intensity plot of the spectroscopic lightcurve residuals for each wavelength bin using the systematic marginalization method. The color bar shows the residuals amplitude for all intensity plots. For the -1 orders we do not compute the transmission below 250\,nm as the flux is too low to produce convergent results in the systematic analysis.  }
    \label{fig:jesus_HRWmarg}
\end{figure*}

While jitter decorelation uses a fixed systematic model plus the jitter files directly from the telescope as a main decorelation factor, systematic marginalization derives its information from evidence obtained from an array of independent systematic models. Systematic marginalization therefore accounts for the unknown factors affecting the lightcurves by weighting them according to the reduced data rather than the telescopes fine guidance sensors. Using systematic marginalization we find that each transit and spectral order favors slightly different combinations of systematic corrections. 
For visit 1 both orders predominantly favor models with a quadratic correction to $\delta_\lambda$, while both orders of visit 2 favor a 3rd order $\phi_{HST}$ correction with additional correction for $\delta_\lambda$. 
Given the similarity in the $\delta_\lambda$ trend for each visit and spectral order, as shown in Fig. \ref{fig:sh}, the more favored correction of the HST breathing in visit 2 suggests that this movement on the detector is likely connected with the thermal effects of the telescope and thus the corrections themselves are interchangeable in this specific case where the structure of the systematic is similar.
For each lightcurve there is a marginal preference to correct for a linear trend in time across the whole visit; however, it is slightly more significant in visit 1. This linear trend across the whole visit has been noted in several other HST timeseries observations \citep[e.g.,][]{deming2013,Sing2015a,kreidberg2014a,wakeford2016}, and is thus likely related to the observatory as a whole rather than a specific instrument.
For each visit and order we show the weighting assigned to each systematic model in the systematic marginalization reduction for the broadband analysis in Fig.\ref{fig:weight}, these model weights are later applied to the spectroscopic lightcurves. The weights shown correspond to the systematic models shown in Table 2 of \citet{wakeford2016}. The structure of this grid is such that it first loops through polynomials correcting for $\delta_\lambda$, followed by added corrections for $\phi_{HST}$ with the second half of the grid (25-49) adding in corrections for $\phi_t$. The overall structure of the computed weights shows that the corrections for $\delta_\lambda$ are the dominant factor given causing the loop every four models.

\begin{figure*}
    \centering
    \includegraphics[width=\textwidth]{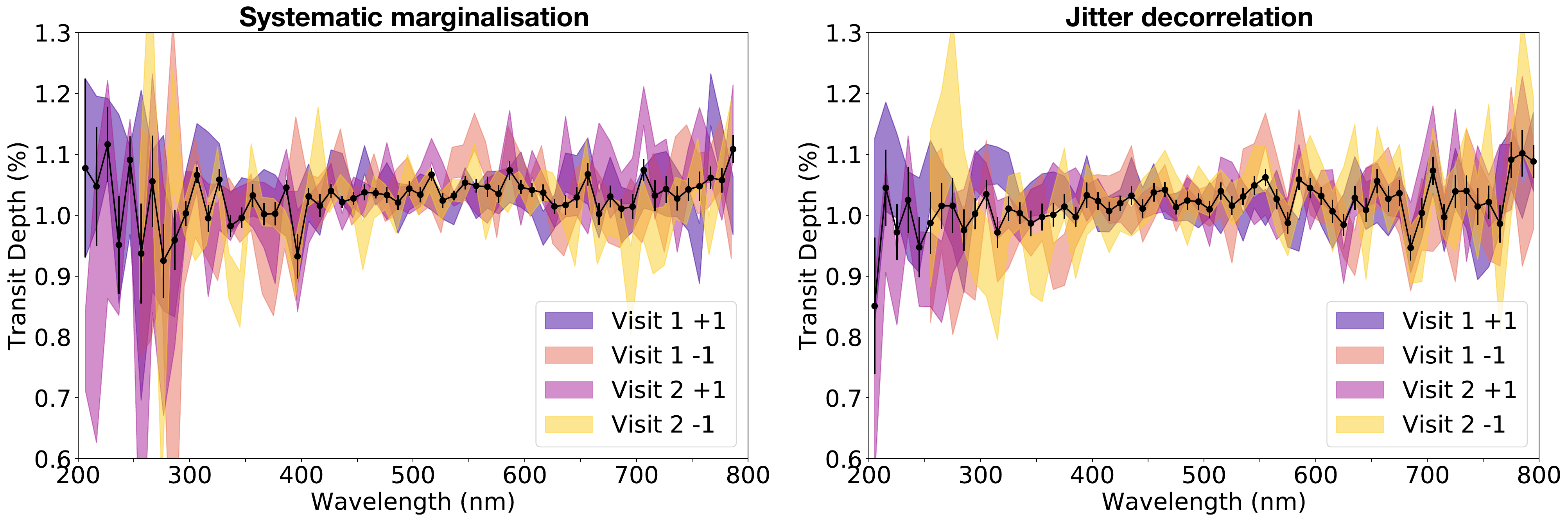}
    \caption{The individual and combined transmission spectra using both systematic marginalization and jitter decorrelation. The two visits and +1/-1 spectral orders are shown as colored shaded regions representing the range of the uncertainties for each spectrum. The final transmission spectrum combining the results of all four are shown as joined black points with errorbars.}
    \label{fig:methods_transmission}
\end{figure*}
% -----------------------
% SECTION:
% Spectroscopic Analysis
% -----------------------
\section{Spectroscopic Analysis}\label{sec:spec_analysis}
To measure the transit depth as a function of wavelength and produce an atmospheric transmission spectrum for HAT-P-41b, we divide the stellar flux into 10\,nm bins ($\sim$5 detector resolution elements) from 200 -- 800\,nm. We note that it is possible to sample the transmission spectrum at a higher resolution ($>$2 resolution elements) in the most optimal portions of the spectrum where the flux is high; however, we use uniform bins across the whole wavelength range for consistency and accurate comparison. 

We analyze each individual spectroscopic lightcurve in the same way, as described in \S\ref{sec:analysis} for the broadband lightcurve, using both systematic marginalization and jitter decorrelation methods. In jitter decorrelation, the systematic correction model is unchanged between wavelength bins, thus assuming all systematics are wavelength independent. Using systematic marginalization, we account for any wavelength dependent systematics by running the full grid of systematic models in each spectroscopic lightcurve. We then use the evidence based weights for each of those models measured in the broadband lightcurve (see Fig\,\ref{fig:weight}) to marginalize over the measured values for each model in each lightcurve. By fixing the systematic model weighting to those derived from the broad-band analysis, the uncertainty is then more representative of the dominant wavelength independent systematics while incorporating the scatter measured across wavelength dependent systematics being fit to the data. 

\begin{figure}
    \centering
    \includegraphics[width=\columnwidth]{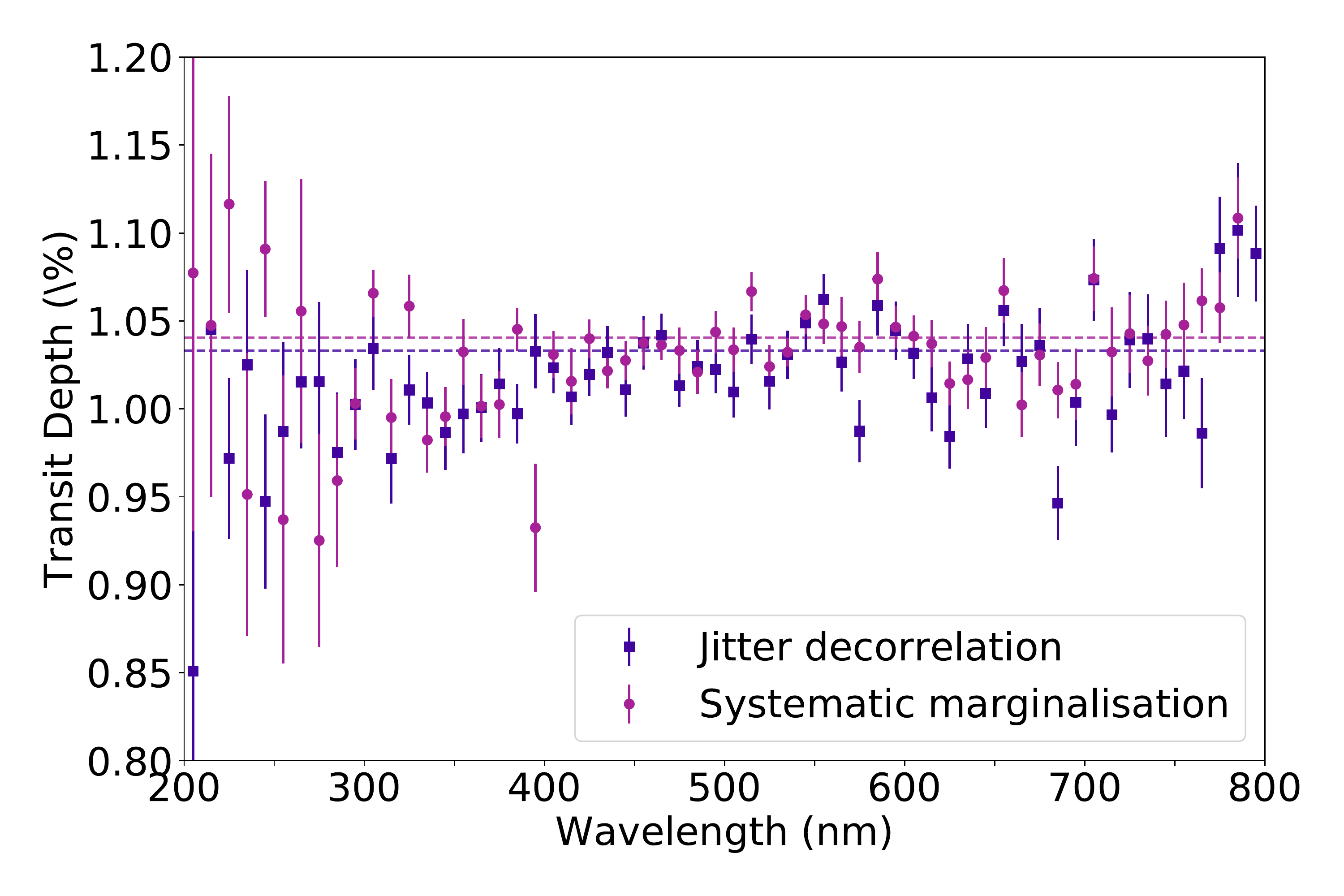}
    \caption{Direct comparison of the final combined transmission spectrum for each systematic treatment: jitter decorrelation (dark squares) and systematic marginalization (light circles). The horizontal dashed lines show the measure broadband depth using each method. }
    \label{fig:comparison_transmission}
\end{figure}

\begin{figure}
    \centering
    \includegraphics[width=\columnwidth]{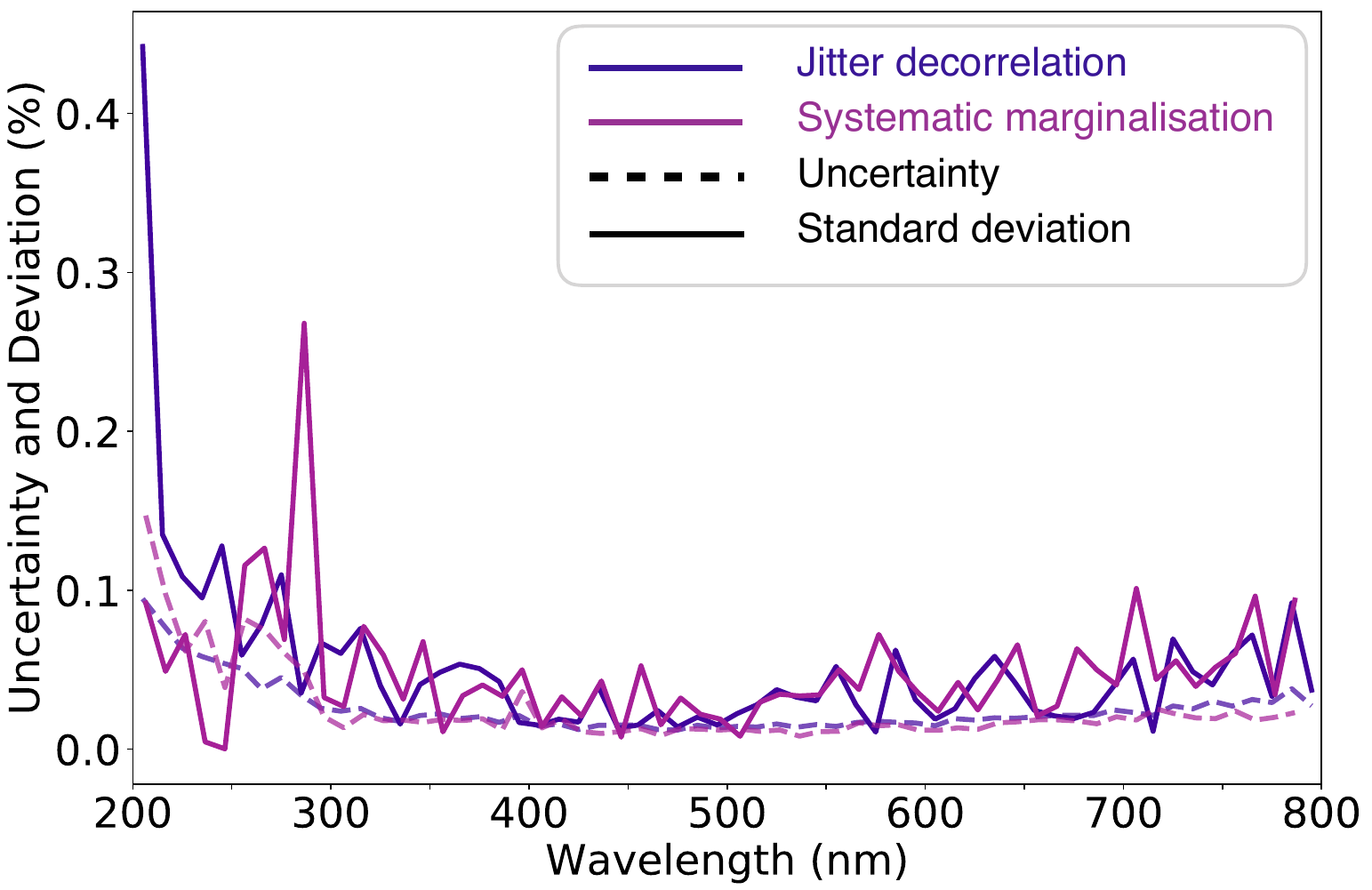}
    \caption{The standard deviation between the four individual transmission spectra in each wavelength bin for systematic marginalization (pink) and jitter decorrelation (purple).}
    \label{fig:comparison_stdev}
\end{figure}

Each visit and +1/-1 spectral orders were analyzed separately using the parameters detailed in Table\,\ref{tab:system_params} fixing the period, inclination, and a/R$_*$, and using the center of transit times listed in Table\,\ref{tab:ephemiris}. Using both jitter decorrelation and systematic marginalization independently, we find consistent results across both visits and spectral orders. Both methods reach photon noise precision in each of the channels determined by calculating the white and red noise associated with the fits (see \citealt{pont2006}), and finding a beta value of 1 consistent with no correlated noise.
We show the residuals from each of the spectroscopic lightcurves for the systematic marginalization analysis in Fig.\,\ref{fig:jesus_HRWmarg} as an intensity residual map to show any global structure in the fit. From the residuals it is clear that the -1 order lightcurves are noisier than the +1 orders. There is also an increase in the scatter at the edges of the wavelength regime, with shorter wavelengths dominating the overall noise range associated with the pure count rates measured from the stellar spectrum in each of the bins (see Fig.\,\ref{fig:stellar_spec}). 

In Fig.\,\ref{fig:methods_transmission}, we present the transmission spectrum measured using both methods for each visit and each +/- first order spectrum with the combined transmission spectrum overlaid. 
We show a direct comparison between the combined transmission spectrum measured using the two systematic treatments in Fig.\,\ref{fig:comparison_transmission}, with 90\% of the points overlapping at the 1-$\sigma$ uncertainty level. 
A direct comparison between the two methods is best demonstrated by looking at the standard deviation and uncertainty in the transit depth measured across the four transits analyzed (see Fig.\,\ref{fig:comparison_stdev}). It is again evident in the standard deviations and uncertainties that the lower counts measured in the near-UV wavelengths ($<$300\,nm) introduce larger scatter and uncertainty to the transit depths. The standard deviation in the short wavelengths indicates that that derived transit depths in each lightcurve are more similar within the uncertainties using systematic marginalization compared to the Jitter decorrelation method. However, there is added scatter with the marginalization method at longer wavelengths. Both methods have similar uncertainty profiles indicating the ability to analyse these data with multiple methods. The unique contribution of the UV points to the transmission spectrum of an exoplanet atmosphere in combination with the optical from a single observation with this low-resolution grism cannot be overstated. 

% -----------------------
% SECTION:
% DISCUSSION
% -----------------------
\section{Discussion}\label{sec:discussion}
We present HST's WFC3/UVIS G280 grism as a reliable observational mode to measure the transmission spectrum of exoplanet atmospheres from 200--800\,nm, critically reaching down to near-UV and optical wavelengths not accessible to JWST. This wavelength range is important to understand and measure cloud opacity sources and their scattering profiles that are defined by the particle sizes (e.g., \citealt{lecavelier2008,wakeford2015, wakeford2017mnras}), escaping atmospheres (e.g., \citealt{ehrenreich2014, sing2019}), and absorption from Na. 

To test this new mode, we measured the atmosphere of the hot Jupiter HAT-P-41b over the course of two consecutive transits with the WFC3/UVIS G280 grism. We obtained the positive and negative first order spectra of the target star in each observation and extracted the stellar flux following the methods outlined by the UVIS calibration pipelines \citep{Kuntschner2009_G280,Rothberg2011_UVISG280,Pirzkal2017_G280}. We analysed the transit data for each visit and spectral order using two well established techniques, instrument systematic marginalization \citep{wakeford2016} and jitter decorrelation \citep{sing2019}. Both analysis techniques produced statistically similar transmission spectra for the atmosphere of HAT-P-41b. We obtain a precision of 29--33\,ppm on the broadband transit depth from 200--800\,nm, and an average precision of $\approx$200\,ppm in 10\,nm spectroscopic bins.

% -----------------------
% Sub-section:
% STIS COMPARISON
% -----------------------
\paragraph{\textbf{Comparison to STIS Observations}}
\begin{figure}
    \centering
    \includegraphics[width=\columnwidth]{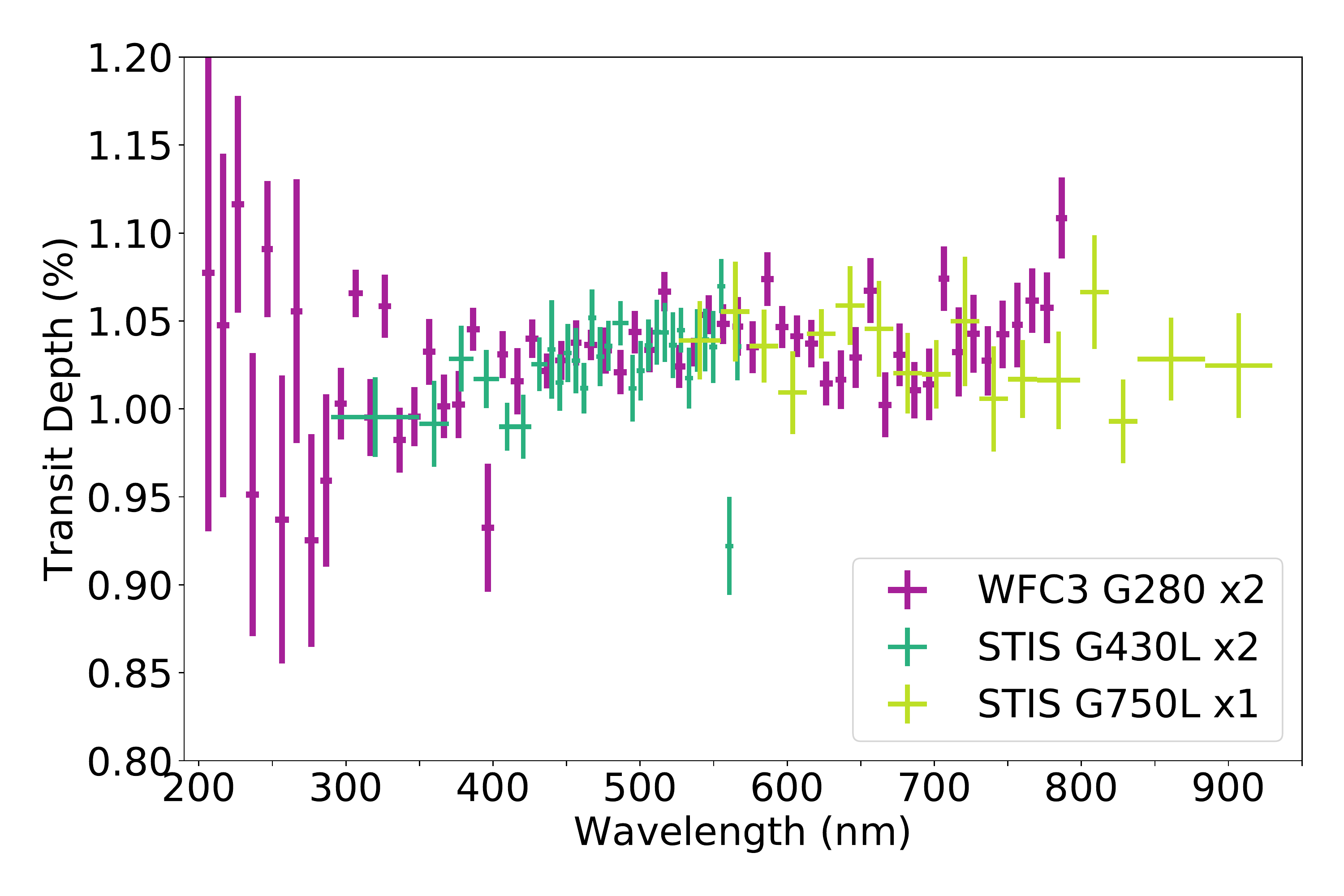}
    \caption{Transmission spectrum of HAT-P-41b measured with WFC3/UVIS G280 grism using systematic marginalization combining two HST observations (pink), compared to STIS G430L combined spectra from two HST observations (dark green) and one observation with the HST STIS G750L grating \citep{sheppard2020}. The WFC3/UVIS G280 grism is able to efficiently measure the atmosphere of a transiting exoplanet from 200--800\,nm to high precision, matching and exceeding that of STIS. }
    \label{fig:comparison_STIS}
\end{figure}

\begin{figure*}[t]
    \centering
    \includegraphics[width=0.95\textwidth]{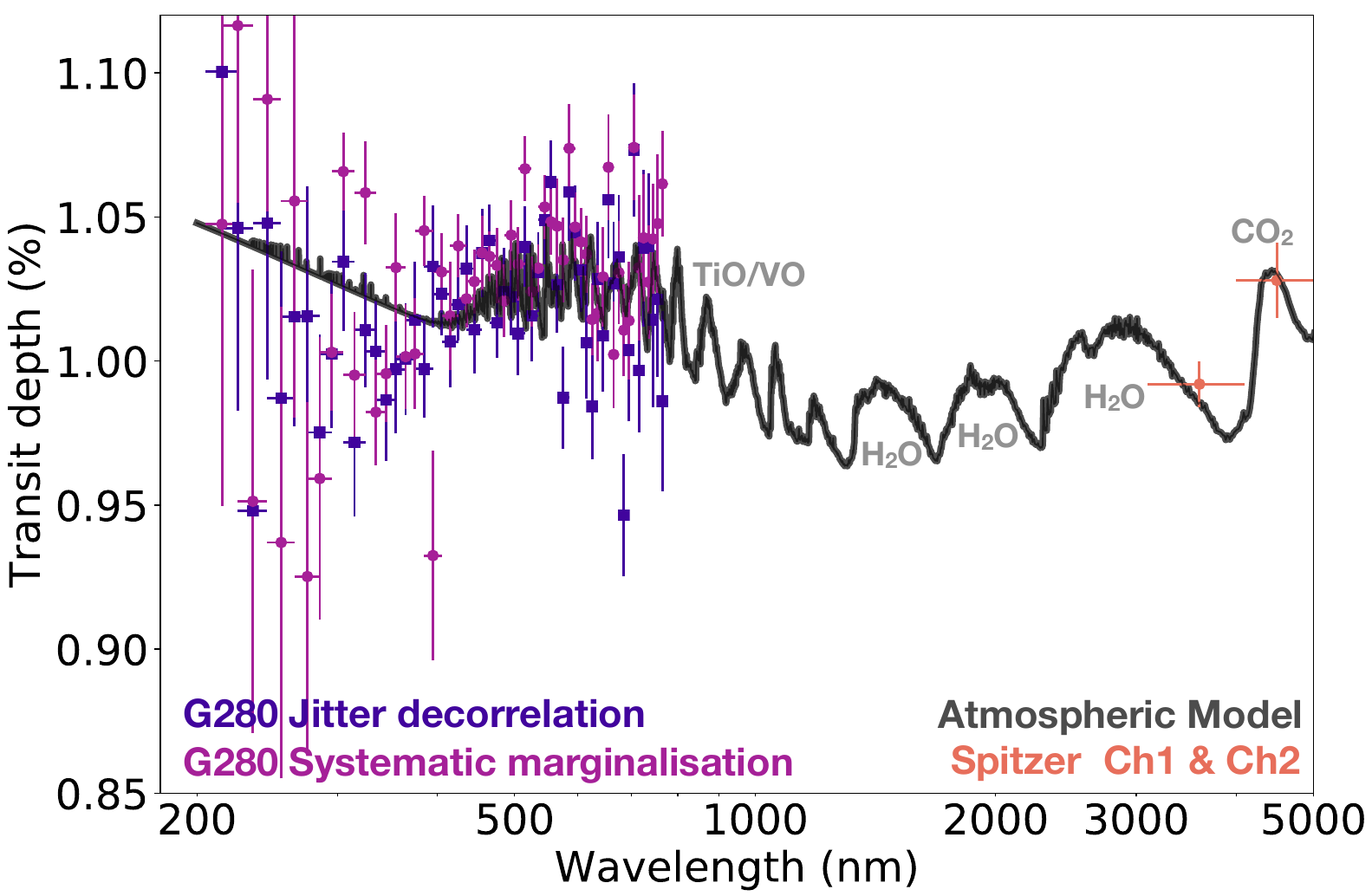}
    \caption{Transmission model fit using the planetary specific grid with rainout condensation by \citet{Goyal2018}. Both the jitter decorrelated and systematic marginalization G280 spectra were fit independently with the \textit{Spitzer} data to the full grid of HAT-P-41b models. Both datasets found the same best fit model with T$_{eq}$\,=\,2091\,K, [M/H]\,=\,+2.0, C/O\,=\,0.7, 1100$\times$scattering, cloud\,=\,0.2. }
    \label{fig:model_fit}
\end{figure*}
We compare the transmission spectrum measured of HAT-P-41b with WFC3/UVIS G280 grism to that measured using STIS G430L and G750L gratings. We find that the combination of the two HST observations in the G280 UVIS grism results in resolution and precision exceeding that of STIS, which required the combination of three HST observations to cover the whole wavelength range compared to two for UVIS. Figure\,\ref{fig:comparison_STIS} shows the transmission spectrum derived using systematic marginalization from two transits with UVIS G280 compared to the transmission spectrum from three transits with STIS G430L and G750L presented by \citet{sheppard2020}. 

Assessing the overall use of UVIS G280 over the STIS gratings, there are a number of trade offs to consider. As G280 cannot be scanned and the throughput is much higher it will likely be more difficult to observe bright (Vmag\,$<$\,7) targets, especially considering the impact of overlapping spectral orders that will make it difficult to extract individual spectral bins at this resolution. Therefore, bright targets will be more efficiently observed with STIS/G430L in particular. Additionally, although UVIS G280 can efficiently measure a wide wavelength range in a single observation it does not extend to wavelengths spanning the potassium absorption line that can only be accurately captured with the STIS G750L grating. However, the extended wavelength coverage into the UV compared to the G430L grism and the comparable resolution means that a potential Na line can be resolved just as easily with UVIS as with STIS but with potentially higher precisions in UVIS. The measured UVIS spectrum far exceeds the resolution and precision over the comparative wavelengths than can be achieved by STIS/G750L (see Fig.\,\ref{fig:comparison_STIS}).

This direct comparison for the same planet demonstrates that the UVIS G280 grism can easily exceed the precision and resolution of STIS in an equivalent number of observations, while being more efficient and requiring less observing time. UVIS G280 also has the advantage of spanning the whole wavelength range in one shot, dramatically reducing the potential impact of stellar activity and systematics which can cause offsets between datasets from different instrument modes. In summary, for targets with Vmag\,$\geqslant$\,7 the UVIS G280 grism shows reduced systematics, higher resolution, precision, and wavelength coverage with more efficient observing compared to STIS G430L and G750L gratings.

% -----------------------
% Sub-section:
% ATMOSPHERIC ESCAPE
% -----------------------
\paragraph{\textbf{Searching for Evidence of Atmospheric Escape}}\label{sec:escape}
The UVIS G280 grism has ideal wavelength coverage to search for signatures of atmospheric escape of the Fe II at 240\,nm and 260\,nm, and the prominent Mg II doublet at 279.63\,nm. A single resolution element for the G280 grism is $\sim$2\,nm which encompasses the whole Mg II doublet absorption line, thus limiting us to strong, low resolution detections. At a single resolution element of the detector, the scatter becomes large and we were unable to converge on a solution to fit the lightcurve systematics. We therefore conducted an analysis of the HAT-P-41b transit data in 4\,nm bins (2 resolution elements) across the 230 -- 290\,nm, with individual moving analyses in 10\,nm steps to search for excess absorption from escaping ions. In this analysis, we find little significant evidence for additional absorption by Fe II and Mg II in the atmosphere. In a single 4\,nm bin centred at 280\,nm we measure additional 0.2\% absorption compared to the average transit depth which could potentially correspond to Mg II. However, this absorption is not seen in bins centered 10\,nm either side of 280\,nm that encompass the peak of the absorption.  The scatter is on the order of 0.3\% across the whole sampled range. 

We conducted our search predominantly using the positive spectral orders for each visit as the throughput and flux levels are high enough for the precision needed at these wavelengths. However, for strong signatures such as those seen in WASP-121b \citep{sing2019} or KELT-9b \citep{Hoeijmakers2018}, which also orbit bright stars, the absorption signature will likely also be measurable in the negative order spectra as well. We conclude that there is no evidence of significant Fe II and Mg II escaping from the atmosphere of HAT-P-41b based on the precision of these measurements. However, we cannot currently conclude where this places HAT-P-41b in the comparative phase space as more measurements with this mode or similar to that shown in \citet{sing2019} will be required over a wide temperature phase space to examine the likelihood of detection.

\paragraph{\textbf{Planetary Specific Model Comparison}}
We ran each of the transmission spectra including the measured \textit{Spitzer} transit depths through the planetary specific forward model grid for HAT-P-41b using rainout condensation presented by \citet{Goyal2018,goyal2019erratum}. In each case, the model fits have the same number of degrees of freedom with the only additional fitting parameter being the absolute altitude of the model. For each UVIS G280 spectrum, we trim the first and last two data points that are likely most affected by low flux and fringing, respectively, and append on the \textit{Spitzer} transit depths. Each transmission spectrum independently favors the same atmospheric model that has: T$_{eq}$\,=\,2091\,K, atmospheric metallicity [M/H]\,=\,+2.0, C/O\,=\,0.7, 1100$\times$scattering profile, and uniform cloud opacity\,=\,0.2 (see Fig.\,\ref{fig:model_fit}).
We find a $\chi^2_\nu$\,=\,1.45 and 1.72 when fitting the most favored model to the jitter decorrelated and marginalized transmission spectrum, respectively.  

The model shows prominent TiO/VO features in the near-UV fitting the UVIS G280 data well in the optical with a wavelength dependent slope associated with a scattering opacity source composed of small sub-micron particles. This model predicts a muted H$_2$O feature in the near-IR that would be detectable with WFC3's G102 and G141 grisms. The \textit{Spitzer} IR is dominated by CO$_2$ that would add additional constraints on the atmospheric metallicity \citep{moses2011} and can be validated by JWST NIRSpec observations. 

% -----------------------
% SECTION:
% CONCLUSIONS
% -----------------------
\section{Conclusions}\label{sec:conclusion}
We present HST's WFC3/UVIS G280 grism as a new and ideal instrument mode for exoplanet time-series characterisation. This is the first time scientific analysis of any observation with this instrument mode has been published.  As such, we provide a detailed breakdown of the challenges and advantages of the instrument, detailed instructions on the spectral extraction with reference to data files and programs provided through UVIS calibration files, and a comparative study of two well established systematic reduction methods. 

To test the UVIS G280 grism for time-series data, we observed the transit of the hot Jupiter HAT-P-41b over two consecutive transit events. This allowed us to measure the overall stability of the instrument, the precision, and resolution without additional concerns associated with potential stellar activity. We obtained both positive and negative first order spectra from each observations providing four different datasets from 200--800\,nm. We analysed each dataset separately before combining the information to produce the final atmospheric transmission spectrum of HAT-P-41b. We applied two different extraction and systematic analysis techniques to the data and find them to be statistically similar across the whole transmission spectrum demonstrating the robust and consistent nature of the instrument critical for accurate exoplanet transmission spectral studies. 

We measure the complete transmission spectrum of the hot Jupiter HAT-P-41b from 200--800\,nm in 10\,nm bins and at 3.6 and 4.5\,$\mu$m with \textit{Spitzer's} IRAC instrument. In the broadband UVIS lightcurves, we reach a precision of 29-33\,ppm, with an average of $\approx$200\,ppm in 10\,nm wide spectroscopic channels. The transmission spectrum shows evidence of TiO/VO in the near-UV to optical with significant absorption from CO$_2$ in the \textit{Spitzer} 4.5\,$\mu$m channel. We fit a grid of forward models specifically derived for HAT-P-41b to the transmission spectrum from multiple reduction pipelines and find constant results with a T$_{eq}$\,=\,2091\,K, [M/H]\,=\,+2.0, C/O\,=\,0.7, scattering $\times$1100, and cloud opacity\,=\,0.2 for rainout condensation (see \citealt{Goyal2018,goyal2019erratum}).
Additional measurements in the near-IR will further aid the interpretation of this planets atmospheric transmission and will be detailed in future publications.

We demonstrate that Hubble's WFC3 UVIS G280 grism is superior to the combination of STIS G430L and G750L gratings for time-series observations in terms of efficiency, precision, and resolution from 300--800\,nm for exoplanet time-series observations. Notably the UVIS G280 grism also allows access to wavelengths as short as 200\,nm with the potential to measure the escaping atmosphere of giant exoplanets via Fe II and Mg II absorption lines and a broad of range of other atmospheric processes. The wavelength coverage offered by the UVIS G280 grism (200--800\,nm) provides a perfect complement to the spectroscopic capabilities of the James Webb Space Telescope (600--14000\,nm), which together can probe the full extent of atmospheric processes in exoplanets that closely orbit their host star.

%\textbf{The UVIS G280 grism is... THE WAY OF THE FUTURE FOR HST and all things!}

% -----------------------
% SECTION:
% Acknowledgements
% -----------------------
\section*{Acknowledgements}
We thank D.\,Deming for use of data from his \textit{Spitzer} program 13044
% \dataset[13044]{https://sha.ipac.caltech.edu/applications/Spitzer/SHA/#id=SearchByProgram&RequestClass=ServerRequest&DoSearch=true&SearchByProgram.field.program=13044&MoreOptions.field.prodtype=aor,pbcd&shortDesc=Program&isBookmarkAble=true&isDrillDownRoot=true&isSearchResult=true} 
that provided the two transits used to obtain accurate system parameters. Thanks to S.\,Morrell for discussions on the stellar parameters with updates from Gaia. We acknowledge private communication from N.\,Nikolov and K.\,Sheppard for the STIS data analysis from the Hubble PanCET program \citep{sheppard2020}.  

This work is based on observations made with the NASA/ESA Hubble Space Telescope, \dataset[HST-GO-15288]{https://archive.stsci.edu/hst/search.php}, that were obtained at the Space Telescope Science Institute, which is operated by the Association of Universities for Research in Astronomy, Inc. A portion of this work is based on observations made with the {\sl Spitzer Space Telescope}, which is operated by the Jet Propulsion Laboratory, California Institute of Technology under a contract with NASA. 
H.R.\,Wakeford acknowledges support from the Giacconi Prize Fellowship at the Space Telescope Science Institute, which is operated by the Association of Universities for Research in Astronomy, Inc.
\software{POET pipeline \citep{stevenson2012,Cubillos2013}, IRAF \citep{tody1986,tody1993}, IDL Astronomy user's library \citep{IDL_ref}, NumPy \citep{numpy}, SciPy \citep{scipy}, MatPlotLib \citep{matplotlib}, AstroPy \citep{astropy}, Photutils \citep{photutils}}
% This analysis made use of components of the IDL astronomy library, IRAF, and the Python packages: NumPy \citep{numpy}, SciPy \citep{scipy}, MatPlotLib \citep{matplotlib}, AstroPy \citep{astropy}, and Photutils \citep{photutils}.
% IRAF is distributed by the{} National Optical Astronomy Observatory, which is operated by the Association of Universities for Research in Astronomy (AURA) under a cooperative agreement with the National Science Foundation.

Author contributions: H.R.\,Wakeford led the UVIS data analysis with detailed comparisons provided by D.K.\,Sing. N.\,Pirzkal provided the UVIS calibration pipeline and knowledge of the instrument. K.B.\,Stevenson analysed the Spitzer data and helped with discussions on the UVIS analysis. N.K.\,Lewis aided with the interpretation of the results and providing context for the observations. T.J.\,Wilson aided in the statistical analysis. All authors provided text and comments for the manuscript. 

% -----------------------
% SECTION:
% REFERENCES
% -----------------------
\bibliography{references}{}

\begin{thebibliography}{}
\expandafter\ifx\csname natexlab\endcsname\relax\def\natexlab#1{#1}\fi
\providecommand{\url}[1]{\href{#1}{#1}}
\providecommand{\dodoi}[1]{doi:~\href{http://doi.org/#1}{\nolinkurl{#1}}}
\providecommand{\doeprint}[1]{\href{http://ascl.net/#1}{\nolinkurl{http://ascl.net/#1}}}
\providecommand{\doarXiv}[1]{\href{https://arxiv.org/abs/#1}{\nolinkurl{https://arxiv.org/abs/#1}}}

\bibitem[{{Astropy Collaboration} {et~al.}(2018){Astropy Collaboration},
  {Price-Whelan}, {Sip{\H{o}}cz}, {G{\"u}nther}, {Lim}, {Crawford}, {Conseil},
  {Shupe}, {Craig}, {Dencheva}, {Ginsburg}, {Vand erPlas}, {Bradley},
  {P{\'e}rez-Su{\'a}rez}, {de Val-Borro}, {Aldcroft}, {Cruz}, {Robitaille},
  {Tollerud}, et~al., \& {Astropy Contributors}}]{astropy}
{Astropy Collaboration}, {Price-Whelan}, A.~M., {Sip{\H{o}}cz}, B.~M., {et~al.}
  2018, \aj, 156, 123, \dodoi{10.3847/1538-3881/aabc4f}

\bibitem[{{Bonomo} {et~al.}(2017){Bonomo}, {Desidera}, {Benatti}, {Borsa},
  {Crespi}, {Damasso}, {Lanza}, {Sozzetti}, {Lodato}, {Marzari}, {Boccato},
  {Claudi}, {Cosentino}, {Covino}, {Gratton}, {Maggio}, {Micela}, {Molinari},
  {Pagano}, {Piotto}, {Poretti}, {Smareglia}, {Affer}, {Biazzo}, {Bignamini},
  {Esposito}, {Giacobbe}, {H{\'e}brard}, {Malavolta}, {Maldonado}, {Mancini},
  {Martinez Fiorenzano}, {Masiero}, {Nascimbeni}, {Pedani}, {Rainer}, \& {Scand
  ariato}}]{bonomo2017}
{Bonomo}, A.~S., {Desidera}, S., {Benatti}, S., {et~al.} 2017, \aap, 602, A107,
  \dodoi{10.1051/0004-6361/201629882}

\bibitem[{{Bourrier} {et~al.}(2018){Bourrier}, {Lecavelier des Etangs},
  {Ehrenreich}, {Sanz-Forcada}, {Allart}, {Ballester}, {Buchhave}, {Cohen},
  {Deming}, {Evans}, {Garcia Munoz}, {Henry}, {Kataria}, {Lavvas}, {Lewis},
  {Lopez-Morales}, {Marley}, {Sing}, \& {Wakeford}}]{Bourrier2018}
{Bourrier}, V., {Lecavelier des Etangs}, A., {Ehrenreich}, D., {et~al.} 2018,
  arXiv e-prints, arXiv:1812.05119.
\newblock \doarXiv{1812.05119}

\bibitem[{Bradley {et~al.}(2019)Bradley, Sipőcz, Robitaille, Tollerud,
  Vinícius, Deil, Barbary, Wilson, Busko, Günther, Cara, Conseil, Droettboom,
  Bostroem, Bray, Bratholm, Lim, Craig, Barentsen, Pascual, Donath, Greco,
  Perren, Kerzendorf, de~Val-Borro, Dencheva, de~Albernaz~Ferreira, Souchereau,
  D'Eugenio, \& Weaver}]{photutils}
Bradley, L., Sipőcz, B., Robitaille, T., {et~al.} 2019, astropy/photutils:
  v0.7.2, v0.7.2,  Zenodo, \dodoi{10.5281/zenodo.3568287}

\bibitem[{{Caswell} {et~al.}(2019){Caswell}, {Droettboom}, {Hunter}, {Firing},
  {Lee}, {Klymak}, {Stansby}, {Sales de Andrade}, {Hedegaard Nielsen},
  {Varoquaux}, {Root}, {Hoffmann}, {Elson}, {May}, {Dale}, {Lee},
  {Sepp{\"a}nen}, {McDougall}, {Straw}, {Hobson}, {Gohlke}, {Yu}, {Ma},
  {Vincent}, {Silvester}, {Moad}, {Katins}, {Kniazev}, {Ariza}, \&
  {Ernest}}]{matplotlib}
{Caswell}, T.~A., {Droettboom}, M., {Hunter}, J., {et~al.} 2019,
  {matplotlib/matplotlib v3.1.0}, v3.1.0,  Zenodo,
  \dodoi{10.5281/zenodo.2893252}

\bibitem[{{Cubillos} {et~al.}(2013){Cubillos}, {Harrington}, {Madhusudhan},
  {Stevenson}, {Hardy}, {Blecic}, {Anderson}, {Hardin}, \&
  {Campo}}]{Cubillos2013}
{Cubillos}, P., {Harrington}, J., {Madhusudhan}, N., {et~al.} 2013, \apj, 768,
  42, \dodoi{10.1088/0004-637X/768/1/42}

\bibitem[{{Cutri} {et~al.}(2012){Cutri}, {Wright}, {Conrow}, {Bauer},
  {Benford}, {Brandenburg}, {Dailey}, {Eisenhardt}, {Evans}, {Fajardo-Acosta},
  {Fowler}, {Gelino}, {Grillmair}, {Harbut}, {Hoffman}, {Jarrett},
  {Kirkpatrick}, {Leisawitz}, {Liu}, {Mainzer}, {Marsh}, {Masci}, {McCallon},
  {Padgett}, {Ressler}, {Royer}, {Skrutskie}, {Stanford}, {Wyatt}, {Tholen},
  {Tsai}, {Wachter}, {Wheelock}, {Yan}, {Alles}, {Beck}, {Grav}, {Masiero},
  {McCollum}, {McGehee}, {Papin}, \& {Wittman}}]{cutri2012_WISE}
{Cutri}, R.~M., {Wright}, E.~L., {Conrow}, T., {et~al.} 2012, {Explanatory
  Supplement to the WISE All-Sky Data Release Products}, Tech. rep.

\bibitem[{{de Wit} {et~al.}(2016){de Wit}, {Wakeford}, {Gillon}, {Lewis},
  {Valenti}, {Demory}, {Burgasser}, {Burdanov}, {Delrez}, {Jehin}, {Lederer},
  {Queloz}, {Triaud}, \& {Van Grootel}}]{dewit2016}
{de Wit}, J., {Wakeford}, H.~R., {Gillon}, M., {et~al.} 2016, \nat, 537, 69,
  \dodoi{10.1038/nature18641}

\bibitem[{Deming {et~al.}(2013)Deming, Wilkins, McCullough, Burrows, Fortney,
  Agol, Dobbs-Dixon, Madhusudhan, Crouzet, Desert, {et~al.}}]{deming2013}
Deming, D., Wilkins, A., McCullough, P., {et~al.} 2013, \apj, 774, 95

\bibitem[{{Eastman} {et~al.}(2010){Eastman}, {Siverd}, \&
  {Gaudi}}]{2010PASP..122..935E}
{Eastman}, J., {Siverd}, R., \& {Gaudi}, B.~S. 2010, \pasp, 122, 935,
  \dodoi{10.1086/655938}

\bibitem[{{Ehrenreich} {et~al.}(2014){Ehrenreich}, {Bonfils}, {Lovis},
  {Delfosse}, {Forveille}, {Mayor}, {Neves}, {Santos}, {Udry}, \&
  {S{\'e}gransan}}]{ehrenreich2014}
{Ehrenreich}, D., {Bonfils}, X., {Lovis}, C., {et~al.} 2014, \aap, 570, A89,
  \dodoi{10.1051/0004-6361/201423809}

\bibitem[{{Evans} {et~al.}(2018){Evans}, {Sing}, {Goyal}, {Nikolov}, {Marley},
  {Zahnle}, {Henry}, {Barstow}, {Alam}, {Sanz-Forcada}, {Kataria}, {Lewis},
  {Lavvas}, {Ballester}, {Ben-Jaffel}, {Blumenthal}, {Bourrier}, {Drummond},
  {Garc{\'\i}a Mu{\~n}oz}, {L{\'o}pez-Morales}, {Tremblin}, {Ehrenreich},
  {Wakeford}, {Buchhave}, {Lecavelier des Etangs}, {H{\'e}brard}, \&
  {Williamson}}]{Evans2018}
{Evans}, T.~M., {Sing}, D.~K., {Goyal}, J.~M., {et~al.} 2018, \aj, 156, 283,
  \dodoi{10.3847/1538-3881/aaebff}

\bibitem[{{Fazio} {et~al.}(2004){Fazio}, {Hora}, {Allen}, {et~al.}}]{IRAC}
{Fazio}, G.~G., {Hora}, J.~L., {Allen}, L.~E., {et~al.} 2004, Astrophy. J.
  Suppl. Ser., 154, 10, \dodoi{10.1086/422843}

\bibitem[{Gelman \& Rubin(1992)}]{Gelman1992}
Gelman, A., \& Rubin, D. 1992, Statistical Science, 7, 457

\bibitem[{{Gilliland} {et~al.}(2010){Gilliland}, {Rajan}, \&
  {Deustua}}]{ISR2010-10}
{Gilliland}, R.~L., {Rajan}, A., \& {Deustua}, S. 2010, {WFC3 UVIS Full Well
  Depths, and Linearity Near and Beyond Saturation}, Tech. rep.

\bibitem[{{Goyal} {et~al.}(2018){Goyal}, {Mayne}, {Sing}, {Drummond},
  {Tremblin}, {Amundsen}, {Evans}, {Carter}, {Spake}, {Baraffe}, {Nikolov},
  {Manners}, {Chabrier}, \& {Hebrard}}]{Goyal2018}
{Goyal}, J.~M., {Mayne}, N., {Sing}, D.~K., {et~al.} 2018, \mnras, 474, 5158,
  \dodoi{10.1093/mnras/stx3015}

\bibitem[{{Goyal} {et~al.}(2019){Goyal}, {Mayne}, {Sing}, {Drummond},
  {Tremblin}, {Amundsen}, {Evans}, {Carter}, {Spake}, {Baraffe}, {Nikolov},
  {Manners}, {Chabrier}, \& {Hebrard}}]{goyal2019erratum}
---. 2019, \mnras, 486, 783, \dodoi{10.1093/mnras/stz755}

\bibitem[{{Hartman} {et~al.}(2012){Hartman}, {Bakos}, {B{\'e}ky}, {Torres},
  {Latham}, {Csubry}, {Penev}, {Shporer}, {Fulton}, {Buchhave}, {Johnson},
  {Howard}, {Marcy}, {Fischer}, {Kov{\'a}cs}, {Noyes}, {Esquerdo}, {Everett},
  {Szklen{\'a}r}, {Quinn}, {Bieryla}, {Knox}, {Hinz}, {Sasselov}, {F{\H
  u}r{\'e}sz}, {Stefanik}, {L{\'a}z{\'a}r}, {Papp}, \&
  {S{\'a}ri}}]{hartman2012}
{Hartman}, J.~D., {Bakos}, G.~{\'A}., {B{\'e}ky}, B., {et~al.} 2012, \aj, 144,
  139, \dodoi{10.1088/0004-6256/144/5/139}

\bibitem[{{Helling}(2019)}]{helling2019}
{Helling}, C. 2019, Annual Review of Earth and Planetary Sciences, 47, 583,
  \dodoi{10.1146/annurev-earth-053018-060401}

\bibitem[{{Hoeijmakers} {et~al.}(2018){Hoeijmakers}, {Ehrenreich}, {Heng},
  {Kitzmann}, {Grimm}, {Allart}, {Deitrick}, {Wyttenbach}, {Oreshenko}, {Pino},
  {Rimmer}, {Molinari}, \& {Di Fabrizio}}]{Hoeijmakers2018}
{Hoeijmakers}, H.~J., {Ehrenreich}, D., {Heng}, K., {et~al.} 2018, \nat, 560,
  453, \dodoi{10.1038/s41586-018-0401-y}

\bibitem[{{Kreidberg} {et~al.}(2014){Kreidberg}, {Bean}, {D{\'e}sert},
  {Benneke}, {Deming}, {Stevenson}, {Seager}, {Berta-Thompson}, {Seifahrt}, \&
  {Homeier}}]{kreidberg2014a}
{Kreidberg}, L., {Bean}, J.~L., {D{\'e}sert}, J.-M., {et~al.} 2014, \nat, 505,
  69, \dodoi{10.1038/nature12888}

\bibitem[{{Kreidberg} {et~al.}(2018){Kreidberg}, {Line}, {Parmentier},
  {Stevenson}, {Louden}, {Bonnefoy}, {Faherty}, {Henry}, {Williamson},
  {Stassun}, {Bean}, {Fortney}, {Showman}, {D{\'e}sert}, \&
  {Arcangeli}}]{kreidberg2018}
{Kreidberg}, L., {Line}, M.~R., {Parmentier}, V., {et~al.} 2018, ArXiv
  e-prints.
\newblock \doarXiv{1805.00029}

\bibitem[{{Kuntschner} {et~al.}(2009){Kuntschner}, {Bushouse}, {K{\"u}mmel}, \&
  {Walsh}}]{Kuntschner2009_G280}
{Kuntschner}, H., {Bushouse}, H., {K{\"u}mmel}, M., \& {Walsh}, J.~R. 2009,
  {The ground calibrations of the WFC3/UVIS G280 grism}, Tech. rep.

\bibitem[{{Landsman}(1995)}]{IDL_ref}
{Landsman}, W.~B. 1995, Astronomical Society of the Pacific Conference Series,
  Vol.~77, {The IDL Astronomy User's Library}, ed. R.~A. {Shaw}, H.~E. {Payne},
  \& J.~J.~E. {Hayes}, 437

\bibitem[{{Lecavelier des Etangs} {et~al.}(2008){Lecavelier des Etangs},
  {Pont}, {Vidal-Madjar}, \& {Sing}}]{lecavelier2008}
{Lecavelier des Etangs}, A., {Pont}, F., {Vidal-Madjar}, A., \& {Sing}, D.
  2008, \aap, 481, L83, \dodoi{10.1051/0004-6361:200809388}

\bibitem[{{Lothringer} {et~al.}(2018){Lothringer}, {Benneke}, {Crossfield},
  {Henry}, {Morley}, {Dragomir}, {Barman}, {Knutson}, {Kempton}, {Fortney},
  {McCullough}, \& {Howard}}]{Lothringer2018b}
{Lothringer}, J.~D., {Benneke}, B., {Crossfield}, I.~J.~M., {et~al.} 2018, \aj,
  155, 66, \dodoi{10.3847/1538-3881/aaa008}

\bibitem[{{Magic} {et~al.}(2015){Magic}, {Chiavassa}, {Collet}, \&
  {Asplund}}]{Magic2015}
{Magic}, Z., {Chiavassa}, A., {Collet}, R., \& {Asplund}, M. 2015, \aap, 573,
  A90, \dodoi{10.1051/0004-6361/201423804}

\bibitem[{{Mandel} \& {Agol}(2002)}]{mandelagol2002}
{Mandel}, K., \& {Agol}, E. 2002, \apjl, 580, L171, \dodoi{10.1086/345520}

\bibitem[{{Marley} {et~al.}(2013){Marley}, {Ackerman}, {Cuzzi}, \&
  {Kitzmann}}]{marley2013}
{Marley}, M.~S., {Ackerman}, A.~S., {Cuzzi}, J.~N., \& {Kitzmann}, D. 2013,
  {Clouds and Hazes in Exoplanet Atmospheres}, ed. S.~J. {Mackwell}, A.~A.
  {Simon-Miller}, J.~W. {Harder}, \& M.~A. {Bullock}, 367--391,
  \dodoi{10.2458/azu_uapress_9780816530595-ch15}

\bibitem[{McCullough \& MacKenty(2012)}]{mccullough2012}
McCullough, P., \& MacKenty, J. 2012, STScI Instrument Science Report WFC3, 8,
  2012

\bibitem[{{McCullough} {et~al.}(2014){McCullough}, {Crouzet}, {Deming}, \&
  {Madhusudhan}}]{mccullough2014}
{McCullough}, P.~R., {Crouzet}, N., {Deming}, D., \& {Madhusudhan}, N. 2014,
  \apj, 791, 55, \dodoi{10.1088/0004-637X/791/1/55}

\bibitem[{{Morrell} \& {Naylor}(2019)}]{morrell2019}
{Morrell}, S., \& {Naylor}, T. 2019, \mnras, 489, 2615,
  \dodoi{10.1093/mnras/stz2242}

\bibitem[{{Moses} {et~al.}(2011){Moses}, {Visscher}, {Fortney}, {Showman},
  {Lewis}, {Griffith}, {Klippenstein}, {Shabram}, {Friedson}, {Marley}, \&
  {Freedman}}]{moses2011}
{Moses}, J.~I., {Visscher}, C., {Fortney}, J.~J., {et~al.} 2011, \apj, 737, 15,
  \dodoi{10.1088/0004-637X/737/1/15}

\bibitem[{{Nikolov} {et~al.}(2014){Nikolov}, {Sing}, {Pont}, {Burrows},
  {Fortney}, {Ballester}, {Evans}, {Huitson}, {Wakeford}, {Wilson}, \& {et
  al.}}]{nikolov2014}
{Nikolov}, N., {Sing}, D.~K., {Pont}, F., {et~al.} 2014, \mnras, 437, 46,
  \dodoi{10.1093/mnras/stt1859}

\bibitem[{Oliphant(2006--)}]{numpy}
Oliphant, T. 2006--, {NumPy}: A guide to {NumPy}, USA: Trelgol Publishing.
\newblock \url{http://www.numpy.org/}

\bibitem[{{Pirzkal} {et~al.}(2017){Pirzkal}, {Hilbert}, \&
  {Rothberg}}]{Pirzkal2017_G280}
{Pirzkal}, N., {Hilbert}, B., \& {Rothberg}, B. 2017, {Trace and Wavelength
  Calibrations of the UVIS G280 +1/-1 Grism Orders}, Tech. rep.

\bibitem[{{Pont} {et~al.}(2006){Pont}, {Zucker}, \& {Queloz}}]{pont2006}
{Pont}, F., {Zucker}, S., \& {Queloz}, D. 2006, \mnras, 373, 231,
  \dodoi{10.1111/j.1365-2966.2006.11012.x}

\bibitem[{{Rothberg} {et~al.}(2011){Rothberg}, {Pirzkal}, \&
  {Baggett}}]{Rothberg2011_UVISG280}
{Rothberg}, B., {Pirzkal}, N., \& {Baggett}, S. 2011, {First Results from
  Contamination Monitoring with the WFC3 UVIS G280 Grism}, Tech. rep.

\bibitem[{{Sheppard}(2020 in prep - private communication)}]{sheppard2020}
{Sheppard}, K. 2020 in prep - private communication, \aj

\bibitem[{{Sing} {et~al.}(2011){Sing}, {Pont}, {Aigrain}, {Charbonneau},
  {D{\'e}sert}, {Gibson}, {Gilliland}, {Hayek}, {Henry}, {Knutson}, {Lecavelier
  Des Etangs}, {Mazeh}, \& {Shporer}}]{sing2011b}
{Sing}, D.~K., {Pont}, F., {Aigrain}, S., {et~al.} 2011, \mnras, 416, 1443,
  \dodoi{10.1111/j.1365-2966.2011.19142.x}

\bibitem[{{Sing} {et~al.}(2015){Sing}, {Wakeford}, {Showman}, {Nikolov},
  {Fortney}, {Burrows}, {Ballester}, {Deming}, \& {et al.}}]{Sing2015a}
{Sing}, D.~K., {Wakeford}, H.~R., {Showman}, A.~P., {et~al.} 2015, \mnras, 446,
  2428, \dodoi{10.1093/mnras/stu2279}

\bibitem[{Sing {et~al.}(2016)Sing, Fortney, Nikolov, Wakeford, Kataria, Evans,
  Aigrain, Ballester, Burrows, Deming, D{\'e}sert, Gibson, Henry, Huitson,
  Knutson, Etangs, Pont, Showman, Vidal-Madjar, Williamson, \&
  Wilson}]{sing2016_nature}
Sing, D.~K., Fortney, J.~J., Nikolov, N., {et~al.} 2016, Nature, 529, 59.
\newblock \url{http://dx.doi.org/10.1038/nature16068}

\bibitem[{{Sing} {et~al.}(2019){Sing}, {Lavvas}, {Ballester}, {Lecavelier des
  Etangs}, {Marley}, {Nikolov}, {Ben-Jaffel}, {Bourrier}, {Buchhave}, {Deming},
  {Ehrenreich}, {Mikal-Evans}, {Kataria}, {Lewis}, {L{\'o}pez-Morales},
  {Garc{\'\i}a Mu{\~n}oz}, {Henry}, {Sanz-Forcada}, {Spake}, {Wakeford}, \&
  {The PanCET collaboration}}]{sing2019}
{Sing}, D.~K., {Lavvas}, P., {Ballester}, G.~E., {et~al.} 2019, \aj, 158, 91,
  \dodoi{10.3847/1538-3881/ab2986}

\bibitem[{{Stevenson} {et~al.}(2012{\natexlab{a}}){Stevenson}, {Harrington},
  {Fortney}, {Loredo}, {et~al.}}]{Stevenson2011}
{Stevenson}, K.~B., {Harrington}, J., {Fortney}, J.~J., {Loredo}, T.~J.,
  {et~al.} 2012{\natexlab{a}}, \apj, 754, 136,
  \dodoi{10.1088/0004-637X/754/2/136}

\bibitem[{{Stevenson} {et~al.}(2012{\natexlab{b}}){Stevenson}, {Harrington},
  {Fortney}, {Loredo}, {Hardy}, {Nymeyer}, {Bowman}, {Cubillos}, {Bowman}, \&
  {Hardin}}]{stevenson2012}
{Stevenson}, K.~B., {Harrington}, J., {Fortney}, J.~J., {et~al.}
  2012{\natexlab{b}}, \apj, 754, 136, \dodoi{10.1088/0004-637X/754/2/136}

\bibitem[{ter Braak \& Vrugt(2008)}]{terBraak2008}
ter Braak, C., \& Vrugt, J. 2008, Statistics and Computing, 18, 435,
  \dodoi{10.1007/s11222-008-9104-9}

\bibitem[{{Tody}(1986)}]{tody1986}
{Tody}, D. 1986, Society of Photo-Optical Instrumentation Engineers (SPIE)
  Conference Series, Vol. 627, {The IRAF Data Reduction and Analysis System},
  ed. D.~L. {Crawford}, 733, \dodoi{10.1117/12.968154}

\bibitem[{{Tody}(1993)}]{tody1993}
---. 1993, Astronomical Society of the Pacific Conference Series, Vol.~52,
  {IRAF in the Nineties}, ed. R.~J. {Hanisch}, R.~J.~V. {Brissenden}, \&
  J.~{Barnes}, 173

\bibitem[{{van Dokkum}(2001)}]{vandokkum2001}
{van Dokkum}, P.~G. 2001, \pasp, 113, 1420, \dodoi{10.1086/323894}

\bibitem[{{Virtanen} {et~al.}(2019){Virtanen}, {Gommers}, {Burovski},
  {Oliphant}, {Cournapeau}, {Weckesser}, {alexbrc}, {Peterson}, {endolith},
  {Mayorov}, {van der Walt}, {Wilson}, {Laxalde}, {Brett}, {Millman}, {Lars},
  {Nelson}, {Haberland}, {eric-jones}, {Polat}, {Larson}, {Kern}, {Moore},
  {Carey}, {Leslie}, {Perktold}, {Reddy}, {Bharti}, {Feng}, \&
  {Vanderplas}}]{scipy}
{Virtanen}, P., {Gommers}, R., {Burovski}, E., {et~al.} 2019, {scipy/scipy:
  SciPy 1.2.1}, v1.2.1,  Zenodo, \dodoi{10.5281/zenodo.2560881}

\bibitem[{Wakeford {et~al.}(2013)Wakeford, Sing, Deming, Gibson, Fortney,
  Burrows, Ballester, Nikolov, Aigrain, Henry, {et~al.}}]{wakeford2013}
Wakeford, H., Sing, D., Deming, D., {et~al.} 2013, \mnras, 435, 3481

\bibitem[{{Wakeford} \& {Sing}(2015)}]{wakeford2015}
{Wakeford}, H.~R., \& {Sing}, D.~K. 2015, \aap, 573, A122,
  \dodoi{10.1051/0004-6361/201424207}

\bibitem[{Wakeford {et~al.}(2016)Wakeford, Sing, Evans, Deming, \&
  Mandell}]{wakeford2016}
Wakeford, H.~R., Sing, D.~K., Evans, T., Deming, D., \& Mandell, A. 2016, The
  Astrophysical Journal, 819, 10.
\newblock \url{http://stacks.iop.org/0004-637X/819/i=1/a=10}

\bibitem[{{Wakeford} {et~al.}(2017){Wakeford}, {Visscher}, {Lewis}, {Kataria},
  {Marley}, {Fortney}, \& {Mandell}}]{wakeford2017mnras}
{Wakeford}, H.~R., {Visscher}, C., {Lewis}, N.~K., {et~al.} 2017, \mnras, 464,
  4247, \dodoi{10.1093/mnras/stw2639}

\bibitem[{{Wakeford} {et~al.}(2019){Wakeford}, {Wilson}, {Stevenson}, \&
  {Lewis}}]{Wakeford2019RNAAS}
{Wakeford}, H.~R., {Wilson}, T.~J., {Stevenson}, K.~B., \& {Lewis}, N.~K. 2019,
  Research Notes of the American Astronomical Society, 3, 7,
  \dodoi{10.3847/2515-5172/aafc63}

\end{thebibliography}
\bibliographystyle{aasjournal}

\begin{longrotatetable}
\begin{deluxetable*}{c|cc|cc|cccc}
\tablecaption{Transmission spectrum of HAT-P-41b based on the combined spectrum of +1 and -1 spectral orders over two transt events. for both systematic marginalization and jitter decorrelation \label{chartable}}
\tablewidth{700pt}
\tabletypesize{\scriptsize}
\tablehead{
 & \multicolumn{2}{c|}{\textbf{Systematic marginalization}} & \multicolumn{2}{c|}{\textbf{Jitter Decorrelation}} & \multicolumn{4}{c}{\textbf{Limb-darkening coefficients$^*$}}\\
\colhead{Wavelength} & \colhead{ Transit depth } & \colhead{Uncertainty} & \colhead{ Transit depth } & \colhead{Uncertainty} & \colhead{c1} & \colhead{c2} & \colhead{c3} & \colhead{c4} \\ 
\colhead{(nm)} & \colhead{(\%)} & \colhead{(\%)} & \colhead{(\%)} & 
\colhead{(\%)} & \colhead{} & \colhead{} &
\colhead{} & \colhead{} 
} 
\startdata
    \textbf{200-800} & \textbf{1.04056} & \textbf{0.00293} & \textbf{1.03303}  &  \textbf{0.00331} & \\
    \hline
       205 &   1.07731 &   0.14700 &   0.85089 &   0.11217 &   0.27642 &  -0.27861 &   0.17437 &   0.81199 \\ 
       215 &   1.04749 &   0.09769 &   1.04495 &   0.06266 &   0.44951 &  -1.31326 &   2.39714 &  -0.53880 \\ 
       225 &   1.11641 &   0.06163 &   0.97186 &   0.04573 &   0.26022 &  -0.34716 &   0.79494 &   0.26854 \\ 
       235 &   0.95137 &   0.08046 &   1.02503 &   0.05374 &   0.44189 &  -0.57725 &   1.16157 &  -0.07411 \\ 
       245 &   1.09086 &   0.03881 &   0.94739 &   0.04950 &   0.54045 &  -1.11394 &   2.37372 &  -0.82917 \\ 
       255 &   0.93706 &   0.08189 &   0.98717 &   0.05082 &   0.48574 &  -0.52757 &   1.52725 &  -0.54102 \\ 
       265 &   1.05554 &   0.07504 &   1.01539 &   0.03792 &   0.33614 &  -0.12161 &   1.23718 &  -0.49873 \\ 
       275 &   0.92522 &   0.06059 &   1.01556 &   0.04516 &   0.51787 &  -0.22060 &   0.71649 &  -0.07533 \\ 
       285 &   0.95923 &   0.04902 &   0.97523 &   0.03401 &   0.47085 &  -0.29824 &   1.30007 &  -0.53323 \\ 
       295 &   1.00307 &   0.02041 &   1.00242 &   0.02578 &   0.43969 &  -0.17729 &   1.45827 &  -0.79410 \\ 
       305 &   1.06577 &   0.01350 &   1.03449 &   0.02395 &   0.40399 &   0.12345 &   0.91689 &  -0.53110 \\ 
       315 &   0.99512 &   0.02188 &   0.97180 &   0.02567 &   0.34789 &   0.38909 &   0.43002 &  -0.26221 \\ 
       325 &   1.05841 &   0.01795 &   1.01076 &   0.01979 &   0.41584 &   0.42379 &   0.33271 &  -0.27752 \\ 
       335 &   0.98227 &   0.01837 &   1.00338 &   0.01746 &   0.32808 &   0.83687 &  -0.47180 &   0.19469 \\ 
       345 &   0.99562 &   0.01689 &   0.98657 &   0.02122 &   0.50090 &   0.47375 &  -0.04764 &  -0.04352 \\ 
       355 &   1.03245 &   0.01882 &   0.99715 &   0.02232 &   0.49485 &   0.49152 &  -0.08977 &  -0.02466 \\ 
       365 &   1.00155 &   0.01805 &   1.00064 &   0.01927 &   0.49130 &   0.55355 &  -0.23093 &   0.04401 \\ 
       375 &   1.00250 &   0.01921 &   1.01414 &   0.02044 &   0.51901 &   0.65211 &  -0.42396 &   0.11740 \\ 
       385 &   1.04525 &   0.01217 &   0.99727 &   0.01692 &   0.44683 &   0.40462 &   0.26623 &  -0.24303 \\ 
       395 &   0.93250 &   0.03636 &   1.03279 &   0.02120 &   0.48246 &   0.16780 &   0.44794 &  -0.21622 \\ 
       405 &   1.03092 &   0.01341 &   1.02334 &   0.01455 &   0.48142 &   0.45253 &   0.03247 &  -0.08316 \\ 
       415 &   1.01568 &   0.01890 &   1.00675 &   0.01593 &   0.43310 &   0.43124 &   0.12149 &  -0.10687 \\ 
       425 &   1.03996 &   0.01103 &   1.01964 &   0.01239 &   0.47865 &   0.31563 &   0.19400 &  -0.11728 \\ 
       435 &   1.02160 &   0.00993 &   1.03207 &   0.01504 &   0.50686 &   0.42439 &  -0.02042 &  -0.06239 \\ 
       445 &   1.02758 &   0.01111 &   1.01087 &   0.01519 &   0.62723 &   0.00768 &   0.49756 &  -0.26869 \\ 
       455 &   1.03755 &   0.01292 &   1.03748 &   0.01523 &   0.66796 &  -0.03296 &   0.39110 &  -0.16936 \\ 
       465 &   1.03632 &   0.00862 &   1.04188 &   0.01242 &   0.64089 &   0.10690 &   0.17586 &  -0.07282 \\ 
       475 &   1.03320 &   0.01313 &   1.01325 &   0.01207 &   0.68519 &   0.04928 &   0.14144 &  -0.03468 \\ 
       485 &   1.02089 &   0.01256 &   1.02408 &   0.01507 &   0.79901 &  -0.19234 &   0.36598 &  -0.16162 \\ 
       495 &   1.04370 &   0.01210 &   1.02236 &   0.01338 &   0.77433 &  -0.20489 &   0.40845 &  -0.15622 \\ 
       505 &   1.03363 &   0.01274 &   1.00958 &   0.01437 &   0.71140 &  -0.05044 &   0.22921 &  -0.08213 \\ 
       515 &   1.06672 &   0.01122 &   1.03961 &   0.01399 &   0.64107 &  -0.00460 &   0.31643 &  -0.15507 \\ 
       525 &   1.02409 &   0.01215 &   1.01568 &   0.01591 &   0.73764 &  -0.16794 &   0.39393 &  -0.17123 \\ 
       535 &   1.03226 &   0.00823 &   1.03075 &   0.01383 &   0.79821 &  -0.32921 &   0.54626 &  -0.23226 \\ 
       545 &   1.05347 &   0.01105 &   1.04898 &   0.01551 &   0.82715 &  -0.38564 &   0.54019 &  -0.20221 \\ 
       555 &   1.04829 &   0.01132 &   1.06219 &   0.01434 &   0.81141 &  -0.35516 &   0.49708 &  -0.18347 \\ 
       565 &   1.04683 &   0.01666 &   1.02657 &   0.01682 &   0.82890 &  -0.41148 &   0.54895 &  -0.20500 \\ 
       575 &   1.03509 &   0.01478 &   0.98730 &   0.01769 &   0.84535 &  -0.43068 &   0.52234 &  -0.18103 \\ 
       585 &   1.07381 &   0.01539 &   1.05872 &   0.01698 &   0.83910 &  -0.44512 &   0.56000 &  -0.20974 \\ 
       595 &   1.04651 &   0.01203 &   1.04453 &   0.01652 &   0.88137 &  -0.54670 &   0.63637 &  -0.22889 \\ 
       605 &   1.04134 &   0.01189 &   1.03169 &   0.01478 &   0.88844 &  -0.57180 &   0.65069 &  -0.23386 \\ 
       615 &   1.03707 &   0.01348 &   1.00632 &   0.01922 &   0.81257 &  -0.38094 &   0.42040 &  -0.13042 \\ 
       625 &   1.01444 &   0.01252 &   0.98434 &   0.01837 &   0.87418 &  -0.59117 &   0.70278 &  -0.27158 \\ 
       635 &   1.01662 &   0.01658 &   1.02845 &   0.01979 &   0.87714 &  -0.58646 &   0.66610 &  -0.24723 \\ 
       645 &   1.02919 &   0.01724 &   1.00875 &   0.01949 &   0.86284 &  -0.56322 &   0.65497 &  -0.25552 \\ 
       655 &   1.06727 &   0.01847 &   1.05593 &   0.02030 &   0.95153 &  -0.74127 &   0.72558 &  -0.26379 \\ 
       665 &   1.00227 &   0.01846 &   1.02700 &   0.02128 &   0.90248 &  -0.67572 &   0.74030 &  -0.27779 \\ 
       675 &   1.03065 &   0.01782 &   1.03617 &   0.02138 &   0.87733 &  -0.62328 &   0.65601 &  -0.22774 \\ 
       685 &   1.01072 &   0.01601 &   0.94646 &   0.02119 &   0.88769 &  -0.66447 &   0.70310 &  -0.25098 \\ 
       695 &   1.01402 &   0.02043 &   1.00378 &   0.02471 &   0.89369 &  -0.70395 &   0.75795 &  -0.27823 \\ 
       705 &   1.07414 &   0.01838 &   1.07324 &   0.02316 &   0.88686 &  -0.68657 &   0.72303 &  -0.26083 \\ 
       715 &   1.03237 &   0.02539 &   0.99668 &   0.02136 &   0.87414 &  -0.65906 &   0.67614 &  -0.23359 \\ 
       725 &   1.04279 &   0.02215 &   1.03922 &   0.02722 &   0.89617 &  -0.72378 &   0.74812 &  -0.26705 \\ 
       735 &   1.02738 &   0.01973 &   1.03978 &   0.02539 &   0.89467 &  -0.73795 &   0.77170 &  -0.28323 \\ 
       745 &   1.04238 &   0.01927 &   1.01427 &   0.03020 &   0.86509 &  -0.64972 &   0.65973 &  -0.23664 \\ 
       755 &   1.04771 &   0.02398 &   1.02146 &   0.02712 &   0.89132 &  -0.75184 &   0.79302 &  -0.29984 \\ 
       765 &   1.06149 &   0.01834 &   0.98611 &   0.03130 &   0.88774 &  -0.75638 &   0.78905 &  -0.29437 \\ 
       775 &   1.05751 &   0.02014 &   1.09123 &   0.02938 &   0.88279 &  -0.73219 &   0.73173 &  -0.25886 \\ 
       785 &   1.10846 &   0.02307 &   1.10165 &   0.03805 &   0.88973 &  -0.77932 &   0.81639 &  -0.31113 \\ 
\multicolumn{9}{l}{* Based on using the +1 order throughput curve and 3D stellar models from \citet{Magic2015}}.\\
\enddata

\end{deluxetable*}
\end{longrotatetable}

\end{document}